\documentclass{article}
\usepackage[final, nonatbib]{neurips_2023_ml4ps}
\usepackage[sort&compress,numbers]{natbib}
\usepackage[utf8]{inputenc} 
\usepackage[T1]{fontenc}   
\usepackage{hyperref}    
\usepackage{url}         
\usepackage{booktabs}    
\usepackage{amsfonts}       
\usepackage{amsmath, amssymb}
\usepackage{nicefrac}  
\usepackage{microtype}
\usepackage{xcolor}
\usepackage{wrapfig} 
\usepackage{aasmacros}  

\usepackage{tikz}
\usepackage{physics}
\usepackage{multicol}
\usepackage{orcidlink}
\usepackage{glossaries}
\hypersetup{
    colorlinks,
    linkcolor={red!50!black},
    citecolor={blue!50!black},
    urlcolor={blue!80!blue}
}

\DeclareRobustCommand{\bbone}{\text{\usefont{U}{bbold}{m}{n}1}}
\DeclareRobustCommand{\bbzero}{\text{\usefont{U}{bbold}{m}{n}0}}

\newacronym{ddpm}{DDPM}{denoising diffusion probabilistic model}
\newacronym{em}{EM}{Euler-Maruyama}
\newacronym{elt}{ELT}{Extremely Large Telescope}
\newacronym{gan}{GAN}{generative adversarial network}
\newacronym{gp}{GP}{Gaussian process}
\newacronym{hst}{HST}{Hubble Space Telescope}
\newacronym{mc}{MC}{Monte Carlo}
\newacronym{pc}{PC}{predictor-corrector}
\newacronym{rim}{RIM}{recurrent inference machine}
\newacronym{sbm}{SBM}{score-based model}
\newacronym{sde}{SDE}{stochastic differential equation}
\newacronym{sie}{SIE}{singular isothermal ellipsoid}
\newacronym{snr}{SNR}{signal-to-noise}
\newacronym{vae}{VAE}{variational autoencoder}
\newacronym{vesde}{VE SDE}{variance-exploding SDE}

\newcommand{\vis}{\mathcal{V}}

\title{Bayesian Imaging for Radio Interferometry with Score-Based Priors}

\author{
    Noé Dia$^{1,2,4}$\orcidlink{0009-0009-6353-0950} \quad 
    M. J. Yantovski-Barth$^{1,2,4}$\orcidlink{0000-0001-5200-4095} \quad 
    Alexandre Adam$^{1,2,4}$\orcidlink{0000-0001-8806-7936} \quad 
    Micah Bowles$^{5}$\orcidlink{0000-0001-5838-8405} \\
    \textbf{Pablo Lemos}$^{1,2,3,4}$ \quad \textbf{Anna M. M. Scaife$^{5,6}$ \quad Yashar Hezaveh$^{1,2,3,4,7,8}$} \\ 
    \textbf{Laurence Perreault-Levasseur$^{1,2,3,4,7,8}$}
    \\
    $^1$Université de Montréal \quad $^2$Ciela Institute \quad $^3$Flatiron Institute \quad $^4$Mila \quad $^5$University of Manchester \\ $^6$The Alan Turing Institute\quad $^7$Trottier Space Institute \quad $^8$Perimeter Institute
    \\
    \texttt{\{noe.dia,michael.barth,alexandre.adam,yashar.hezaveh,}\\ \texttt{laurence.perreault.levasseur\}@umontreal.ca}\\
    \texttt{pablo.lemos@mila.quebec}\\
    \texttt{micah.bowles@postgrad.manchester.ac.uk}\\
    \texttt{anna.scaife@manchester.ac.uk}
}

\begin{document}

\maketitle

\begin{abstract}
The inverse imaging task in radio interferometry is a key limiting factor to retrieving Bayesian uncertainties in radio astronomy in a computationally effective manner. We use a score-based prior derived from optical images of galaxies to recover images of protoplanetary disks from the DSHARP survey. We demonstrate that our method produces plausible posterior samples despite the misspecified galaxy prior. We show that our approach produces results which are competitive with existing radio interferometry imaging algorithms.
\end{abstract}

\section{Introduction}
Interferometry, a powerful technique in astronomy, combines signals from an array of radio antennae to achieve resolutions unattainable by any individual telescope. ALMA, a modern radio interferometer, is capable of high angular resolutions with high sensitivity, which enables the study of objects such as protoplanetary disks \citep[e.g.,][]{HTLau2014} or high-redshift galaxies \citep[e.g.,][]{Laporte2021}. The analysis of interferometric data, however, is exceedingly challenging due to the incomplete sampling of the Fourier space (uv-space) where data is measured. The most accurate methods fit the data directly in uv-space but due to the large number of the measured Fourier modes (visibilities), these methods require extremely expensive computations. A popular avenue is therefore to convert interferometric data to images of the sky.

The most widely-used imaging methods are derivatives of CLEAN \citep{hogbom1974aperture, cornwell2008multiscale, rau2011multi, offringa2017optimized,2009_clean_review}, a deconvolution algorithm which attempts to iteratively remove from the images the effects of the incomplete uv-space sampling. Unfortunately, all versions of CLEAN suffer from several drawbacks: the algorithm can produce images with unphysical negative flux, results in suboptimal resolution, and does not output statistical uncertainties \citep{2021_bayesian_vs_clean}. These limitations have motivated efforts to develop alternative imaging algorithms. One of these approaches is MPoL \citep{mpol}, which adopts a regularized maximum likelihood approach. Still, this technique does not provide uncertainty quantification and relies on fine-tuning of regularization parameters (an implicit prior). Another approach is \texttt{resolve} \citep{2016A&A...586A..76J, 2019A&A...627A.134A, 8553533}, which adopts a Gaussian prior in a Bayesian inference setting. To handle the computational challenges of Bayesian inference, \texttt{resolve} employs Metric Gaussian Variational Inference \citep{knollmüller2020metric}.

In this work, we use score-based generative models \citep{Ho2020,Song2021} as principled priors \citep{graikos2022diffusion} to perform Bayesian inference for radio interferometric imaging. Given a prior (in our case, a trained denoising score network), our formalism is free of hyperparameter tuning and can be adapted on the fly to new datasets without modification. Furthermore, only a fixed number of steps are required to generate samples from a good approximation to the high-dimensional posterior. We test our pipeline on ALMA observations of 3 DSHARP protoplanetary disks \citep{Andrews2018DSHARP} as first application of this method to radio interferometry data. For our prior model, we use high-resolution optical images of galaxies as proxies of images of protoplanetary disks, which would otherwise only be available through simulations \citep[e,g.,][]{FARGO3D}. We show that, even with a misaligned prior, this framework is capable of competing with state-of-the-art methods

\section{Methods}
A radio interferometer measures the visibilities of the sky emission. For distant sources, the van Cittert-Zernike theorem \citep{vancittert1934, zernike1938} states that the visibilities correspond to the Fourier components of the true sky emission. However, since only a subset of the visibilties is observed and since the visibilties are subject to measurement errors (noise), recovering the sky emission is an ill-posed inverse problem.
As such, our goal is to recover plausible surface brightness profiles over a pixel grid, $\mathbf{x} \in \mathbb{R}^n$, where $n$ is the number of pixels of our model, using ALMA observations of a protoplanetary disk, $\vis \in \mathbb{C}^{m}$, where $m$ is the number of gridded visibilities (we expand on the gridding process in the next section).
Measurements are linearly related to sky emission via the equation
\begin{equation}\label{eq:vis}
\vis = S \mathcal{F} P_\text{beam} \mathbf{x} + \boldsymbol{\eta}\, ,
\end{equation}
where $\mathcal{F} \in \mathbb{C}^{n \times n}$ is the dense, unitary, 2D Fourier operator and $S \in \{0,1\}^{m \times n}$ is a mask, otherwise known as the sampling function, which selects visibilities corresponding to measurements. $P_{\mathrm{beam}}$ is the primary beam --- the response function of the primary antenna --- and $\boldsymbol{\eta}$ is additive noise.

In a Bayesian inference setting, solving this problem translates into the task of sampling from the posterior defined by Bayes's theorem as the product of the likelihood, $p(\vis \mid \mathbf{x}) = p(\boldsymbol{\eta})$, which encodes the properties of the additive noise distribution, and the prior, $p(\mathbf{x})$:
\begin{equation}\label{eq:Bayes}
    p(\mathbf{x} \mid \vis) \propto p(\vis \mid \mathbf{x}) p(\mathbf{x})\, .
\end{equation}
The prior represents the known information about the uncertain parameters, here the sky emission, $\mathbf{x}$, before considering any measurement. Although it plays an important role in the statistical inversion, \citet{Feng2023a} show that misaligned score-based priors can recover plausible solutions for the black hole shadow from radio interferometric measurements. This is consistent with a Bayesian inference perspective, in which our knowledge of the world is iteratively updated with new data. When new data is informative, --- i.e., when the likelihood is concentrated or narrow --- and the prior is permissive --- i.e., does not rule out unlikely events --- the posterior can be very different from prior expectations. 

In this work, we apply a score-based model trained on images of galaxies to posterior inference over real ALMA observations of DSHARP protoplanetary disks \citep{Andrews2018DSHARP}. The likelihood of these observations is quite constraining since the number of observed visibilities is large ($m > 10^6$), so it constitutes an ideal test for our hypothesis that a misaligned prior should still allow us to recover plausible solutions. We summarize herein our methodology to characterize the prior and sample from the posterior.

\subsection{Sampling from the posterior with score-based priors}
To characterize the prior, $p(\mathbf{x})$ --- or more specifically its score, $\grad_{\mathbf{x}}\log p(\mathbf{x})$ --- we employ denoising score matching \citep{Hyvarinen2005,Vincent2011,Alain2014}. In summary, we train a neural network based on the U-net architecture \citep{Ronneberger2015} with a weighted Fisher divergence loss \citep{Song2022ml} to match a Gaussian perturbation kernel $p(\mathbf{x}_t \mid \mathbf{x}) = \mathcal{N}(\mu(t) \mathbf{x}, \sigma^2(t) \bbone_{n\times n})$ associated with a Stochastic Differential Equations (SDE), where both the network and the kernel are parameterized by the time index $t \in [0, 1]$ of the SDE \citep{Song2021}. The scalar functions $\mu(t)$ and $\sigma(t)$ are hyper-parameters to specify. In this work, we use the Variance Preserving (VP) SDE with $\beta_{\min} = 0.01$, $\beta_{\max} = 20$ and $\epsilon=10^{-5}$ \citep{Karras2022}. 

The SDE used to sample from the posterior is obtained with Anderson's reverse-time formula 
\citep{Anderson1982} and setting the posterior as the $t=0$ boundary condition of the SDE 
\begin{equation}\label{eq:sde}
    d \mathbf{x} = \big(f(\mathbf{x}, t) - g^{2}(t) \grad_{\mathbf{x}} \log p_t( \mathbf{x} \mid \mathbf{y})\big)dt + g(t) d \bar{\mathbf{w}}\, .
\end{equation}
$\bar{\mathbf{w}}$ is a time-reversed Wiener process and $\mathbf{y} \equiv \vis$ is the observation.
The drift, $f(\mathbf{x}, t)$, and the homogeneous diffusion coefficient, $g(t)$, are specified by the SDE chosen for the prior \citep{Song2021}. Applying Bayes's theorem, we can decompose the posterior score into a prior score and a likelihood score
\begin{equation}\label{eq:score_bayes}
    \grad_\mathbf{x} \log p_t(\mathbf{x} \mid \mathbf{y}) = \grad_{\mathbf{x}} \log p_t (\mathbf{x}) + \grad_\mathbf{x} \log p_t(\mathbf{y} \mid \mathbf{x})
\end{equation} 
Unfortunately, the likelihood score, $\grad_\mathbf{x} \log p_t(\mathbf{y} \mid \mathbf{x})$, is an intractable quantity \citep{Chung2022,Feng2023a,Feng2023b}. 
However, since the likelihoods considered in this work are very informative, we can employ the \textit{convolved likelihood approximation} \citep{Remy2022,Adam2022PosteriorSO}, which allows us to write
\begin{equation}\label{eq:convolved_likelihood}
    p_t(\mathbf{y} \mid \mathbf{x}_t) \approx \mathcal{N}\left(\mu(t)\mathbf{y} \mid A\mathbf{x}_t, \mu^2(t)\Gamma + 
    \sigma^2(t)\Sigma\right)\, ,
\end{equation}
where $A \equiv S \mathcal{F} P_{\mathrm{beam}}$, $\Sigma_{jk} = \frac{1}{2}\delta_{jk} + \frac{1}{2} \delta_{j0}\delta_{k0}$, where $\delta_{jk}$ denotes the Kronecker delta, and $\Gamma \in \mathbb{R}^{2m \times 2m}$ is the covariance matrix of the noise in the gridded visibilities. Note that the complex distribution in equation \eqref{eq:convolved_likelihood} is written in terms of a \textit{vectorized} complex random variable, i.e., 
$\boldsymbol{\eta} \mapsto (\Re(\boldsymbol{\eta}), \Im(\boldsymbol{\eta})) \in \mathbb{R}^{2m}$. 
We refer to the appendix \ref{sec:approx} for more details.

\begin{figure}[t]
    \centering
    \includegraphics[scale = 0.4]{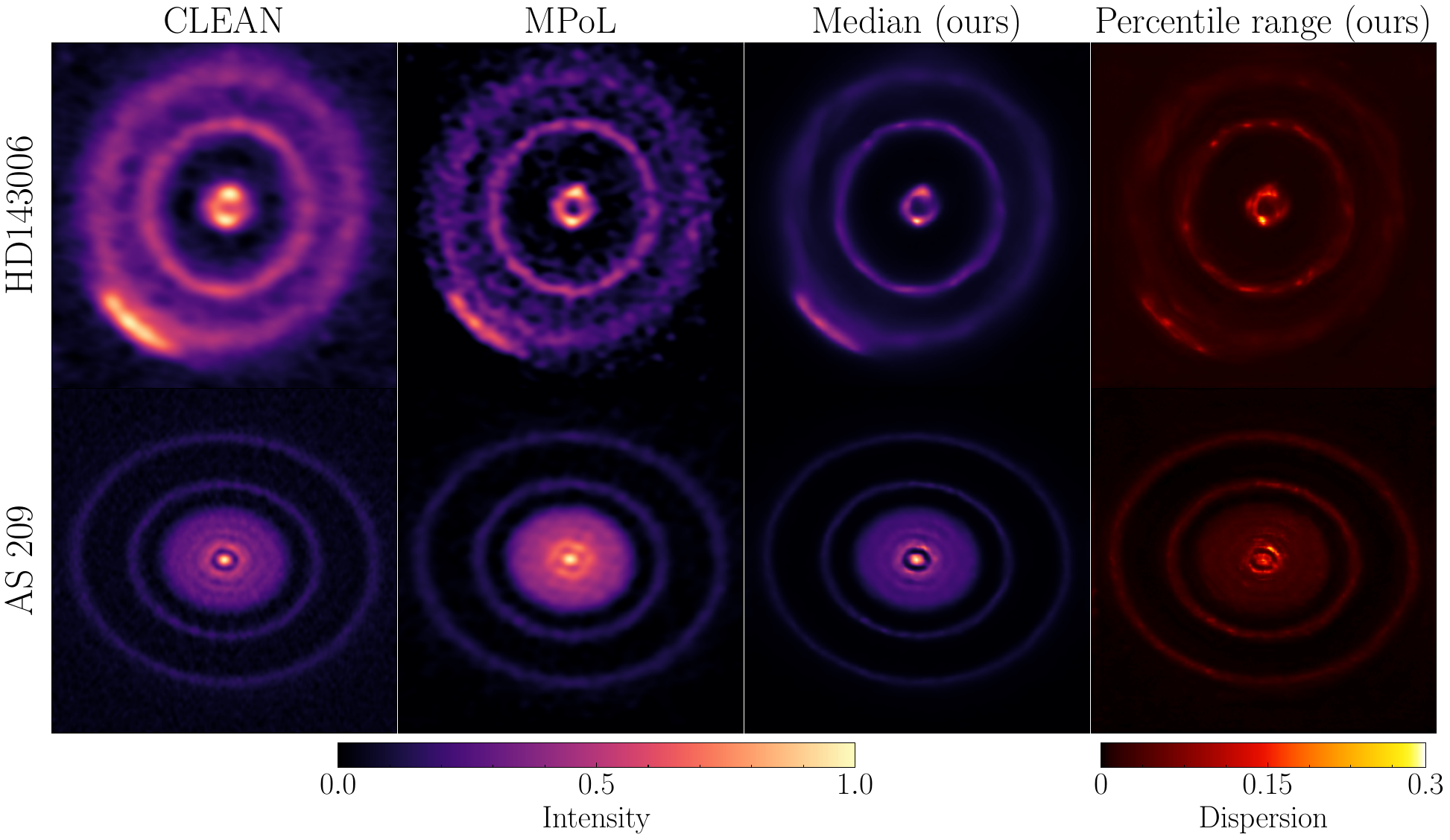}
    \caption{Image reconstructions by CLEAN, MPoL, and our score-based model for two protoplanetary disks. The two columns at right are pixel-wise statistics computed using the posterior obtained from our score-based approach using the SKIRT prior (we also show posterior samples of these two disks in appendix \ref{app_samples}). From left to right the statistics include the median and the 84\%-16\% percentile range. The MPoL image for AS 209 (bottom, second from left) is not fully optimized and is shown purely for illustration.}
    \label{fig_res}
    \vspace{-0.5cm} 
\end{figure}

\section{Data and physical model} \label{sec:data_and_phys}
The \textbf{PROBES} dataset is a compendium of high-quality local late-type galaxies \citep{Stone2019,Stone2021} that we leverage as a prior for our reconstructions. These galaxies, used in previous studies to train diffusion models \citep{Smith2022,Adam2022PosteriorSO}, have resolved structures of spiral arms, bulges, and disks, which can vaguely resemble the structure of protoplanetary disks. 
The \textbf{SKIRT TNG} \citep{Bottrell2023} dataset is a large public collection of images spanning $0.3$-$5$ microns made by applying dust radiative transfer post-processing \citep{Camps2020} to galaxies from the TNG cosmological magneto-hydrodynamical simulations 
\citep{Nelson2019}. The SKIRT TNG dataset offers us an opportunity to compare prior distributions based on the same type of objects --- galaxies --- but with different underlying assumptions about the physics that govern the formation of their structure \citep{Weinberger2017,Pillepich2018}, which in the case of SKIRT TNG is inherited from simulations instead of observations. In this work, we use the $z$ band in both datasets to train the prior score models.

Disk Substructures at High Angular Resolution Project (\textbf{DSHARP}) \citep{Andrews2018DSHARP} is a recent survey aiming to characterize the substructures of 20 nearby protoplanetary disks by observing their continuum emission around 240 GHz with ALMA. High angular resolution observations of such systems can provide a better understanding of the physical processes occurring in a circumstellar disk which ultimately lead to planetary formation.

We process the calibrated \textbf{visibilities} released by the DSHARP survey using the \texttt{visread} package created by the MPoL team \citep{Zawadzki_2023}. For each spectral window, we average the data across the different polarizations and conserve the non-flagged data. We assume a flat spectral index, which permits us to combine spectral windows. In order to perform inference over regular grid models, we bin the visibilities using our implementation of a sinc convolutional gridding function, which was chosen for its uniform effect in image space. We estimate the diagonal part of the covariance matrix, $\Gamma \in \mathbb{R}^{2m \times 2m}$, by computing a weighted standard deviation of the real and imaginary parts of the observed visibilities, $\mathcal{V}$, within each bin. If a bin contains a small number of visibilities, we simultaneously expand the bin and the sinc window function until each cell has at least 5 data points.

In a statistical inference setting, the \textbf{forward model} attempts to capture the physics connecting the parameters of interest (in our case, pixelated sky emission) to the observed data.
For our forward model, we represent equation \eqref{eq:vis} using several simplifying assumptions. We define our images on a regular grid with a constant pixel size and we pad our posterior samples with noise during the diffusion in order to match the resolution of the gridded visibilities in Fourier space. We also model $\mathcal{F}$ with a Fast Fourier Transform (FFT) \citep{CooleyFFT1965}, $P_{\mathrm{beam}}$ with a flat response (i.e., $P_{\mathrm{beam}} \equiv \bbone_{n \times n}$), and the noise, $\boldsymbol{\eta} \in \mathbb{R}^{2m}$, by a Gaussian distribution with diagonal covariance, $p(\boldsymbol{\eta}) = \mathcal{N}(0, \Gamma)$. 

\section{Results and discussion}\label{sec:results}

\begin{figure}[t]
  \centering
  \begin{tikzpicture}
    \node at (4,0) {\includegraphics[scale = 0.35]{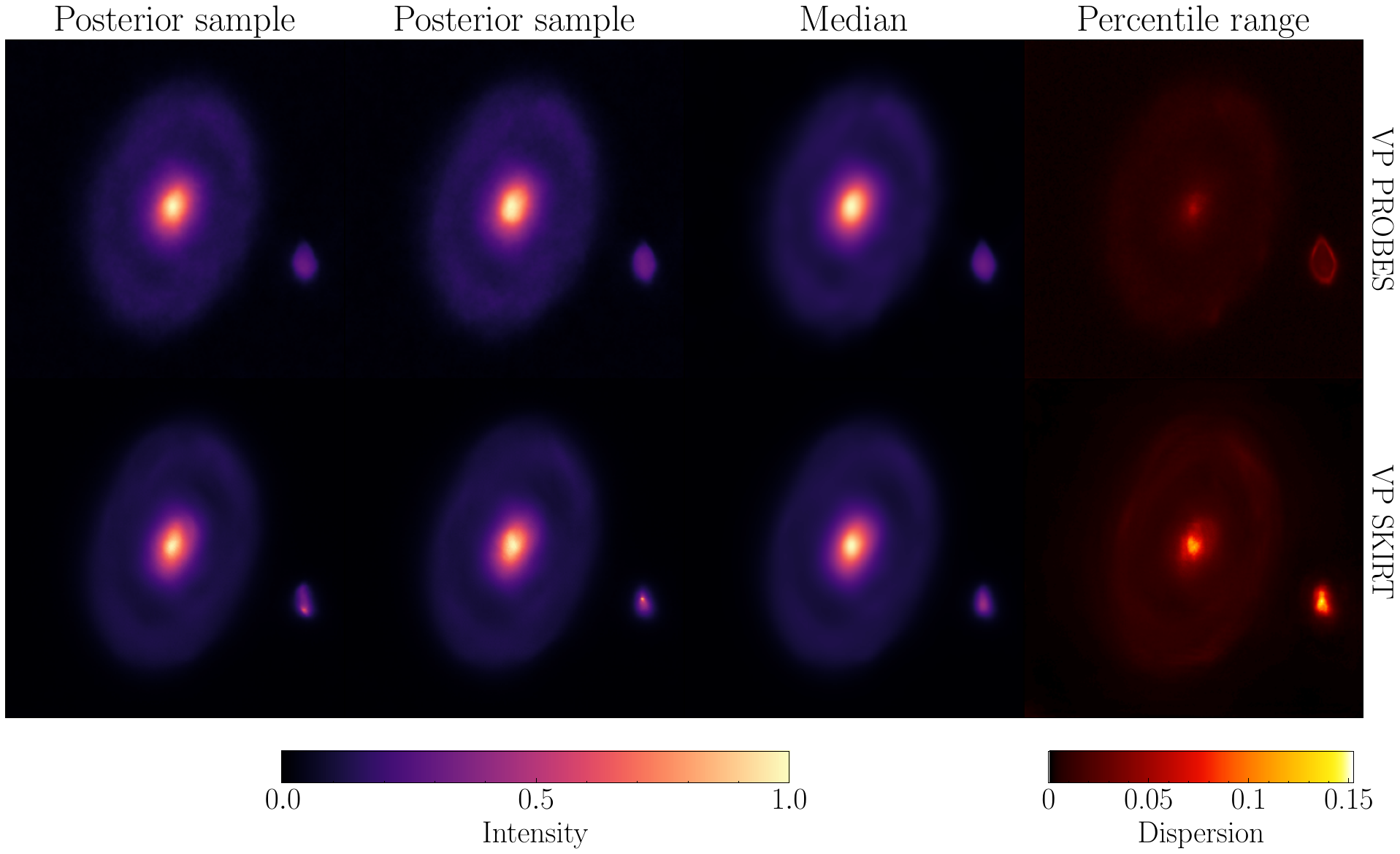}}; 
    
    \node at (-3,0.54) {\includegraphics[scale = 0.3505]
    {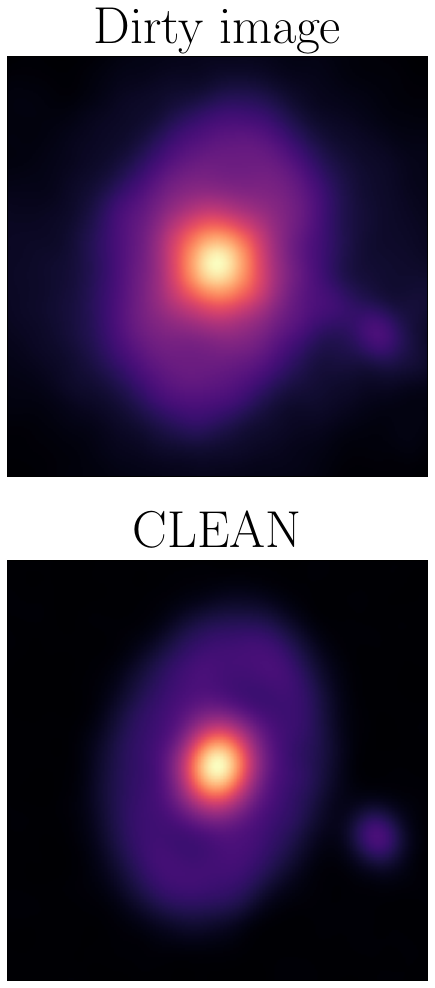}};
  \end{tikzpicture}
  \caption{Image reconstruction of HTLup for two different priors. The first column at left shows from top to bottom the dirty image obtained by "naïvely" applying an inverse Fourier transform on the gridded visibilities and the CLEAN reconstruction published by the DSHARP team~\cite{2018ApJ...869L..44K}. The rest of the figure are posterior samples and pixel-wise statistics (median and 84\%-16\% percentile range) from our approach for a prior trained on two different datasets (PROBES on the first row and SKIRT on the second row). }
  \label{fig_vp}
  \vspace{-0.5cm}
\end{figure}

We test our approach on three protoplanetary disks, HD143006, AS 209, and HT Lup \citep{Andrews2018DSHARP, Zhang_2018} choosing a pixel size of 4 mas, 9 mas and 1.5 mas respectively. 
Fig.~\ref{fig_res} shows the resulting images generated by CLEAN, MPoL, and this work for HD143006 and AS 209. 

The CLEAN images were obtained directly from the DSHARP collaboration \citep{Andrews2018DSHARP}. For the MPoL image of HD143006, we adopted the hyperparameters from the MPoL documentation's tutorial, which were optimized for this disk; we then performed a limited hyperparameter search for the disk AS 209. Thus, the performance of MPoL on the disk AS 209 is not optimized and is shown only for illustrative purposes. For generating our posterior samples --- and since our method is fully parallel --- we use 20 V100 GPUs to get 360 posterior samples ($256 \times 256$ pixels) in $\sim 1$ hour wall-time. Sampling is performed with 4000 steps of the discretized Euler-Maruyama SDE solver.

In order to avoid differences in unit conventions between the different algorithms, we normalize each image's maximum pixel value to 1. Our method shows competitive resolution and dynamic range performance compared to MPoL and CLEAN. 

\begin{wrapfigure}{r}{0.5\textwidth} 
    \centering
    \vspace{-0.25cm}
    \includegraphics[scale=0.45]{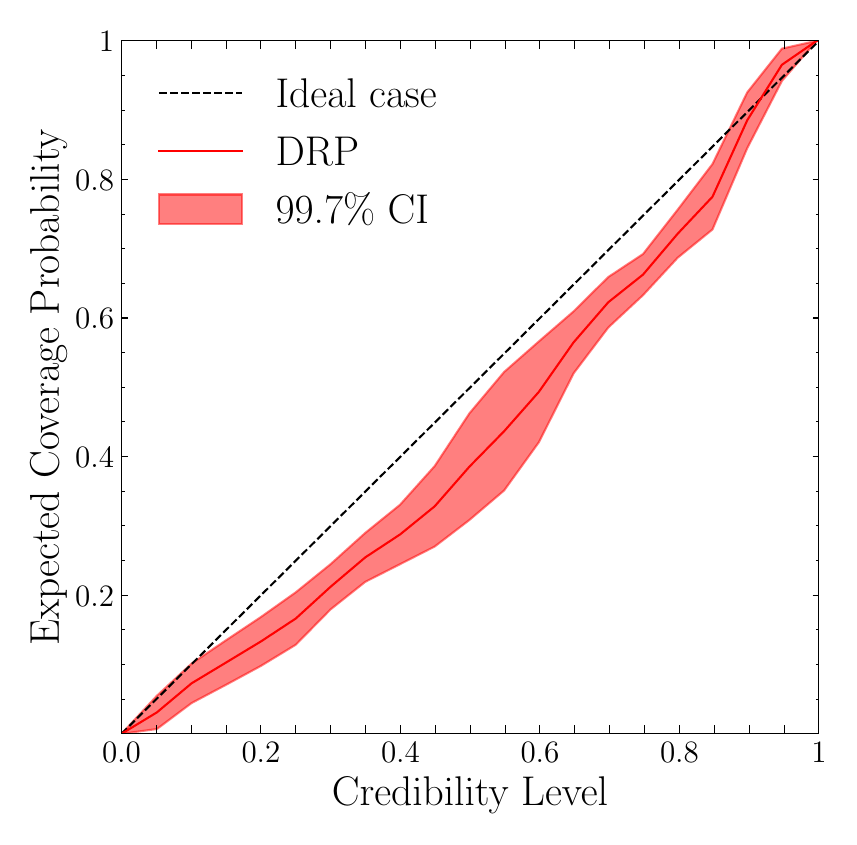}
    \caption{Results from the TARP coverage test for samples generated by our model. The deviation of the red line (this work) at greater than 3-sigma significance from the diagonal indicates a bias in our uncertainty quantification.}
    \label{fig_tarp}
    \vspace{-0.5cm}
\end{wrapfigure}
Next, for HT Lup, we show in Fig.~\ref{fig_vp} the effect of two different priors trained under same SDE but on different datasets. Even though the prior has a clear effect on the posterior's sample statistics, we still get consistent imaging results across the two priors since the likelihood is highly informative and the priors are sufficiently permissive.

Finally, we attempt to assess the coverage of our posterior estimator using Tests of Accuracy with Random Points (TARP) \citep{lemos2023samplingbased}, a method to assess the accuracy of the posterior encoded by a generative model using samples by the model. To speed up the computation of this coverage test, we use a prior trained on a downsampled version of the SKIRT dataset. Our model shows statistically significant deviation from the ideal case (Figure \ref{fig_tarp}), which is a general indicator of a bias in our uncertainty quantification. We believe that the source of this bias may stem from the convolved likelihood approximation. To explore this further, we plan to conduct additional coverage testing of our inference process while varying the built-in assumptions of our modeling. This exploration may uncover potential strategies to mitigate the bias in our approach.

In conclusion, our score-based imaging algorithm's results are shown to be competitive with existing radio interferometry imaging algorithms. Our method, while currently biased, has the potential to provide realistic uncertainty quantification. In future work, we wish to explore the possibility of adding samples of protoplanetary disks from simulations as a way to encode a more accurate prior and quantify the robustness of our reconstructions to prior misspecification.

\section{Acknowledgments}
This research was made possible by a generous donation by Eric and Wendy Schmidt with the recommendation of the Schmidt Futures Foundation. We are grateful for useful discussions with Antoine Bourdin. The work is in part supported by computational resources provided by Calcul Quebec and the Digital Research Alliance of Canada. Y.H. and L.P. acknowledge support from the National Sciences and Engineering Research Council (NSERC) of Canada grant RGPIN-2020-05073 and 05102, the Fonds de recherche du Québec grant 2022-NC-301305 and 300397, and the Canada Research Chairs Program. The work of A.A. was partially funded by an NSERC CGS D scholarship. 
N.D. was supported through an Undergraduate Student Research Award provided by NSERC. 
M.B. \& A.M.M.S. gratefully acknowledge support from the UK Alan Turing Institute under grant reference EP/V030302/1.

Software used: \texttt{astropy} \citep{astropy:2013,astropy:2018}, \texttt{jupyter} \citep{jupyter}, \texttt{matplotlib} \citep{matplotlib} , \texttt{numpy} \citep{numpy}, \texttt{PyTorch} 
 \citep{pytorch}, \texttt{tqdm} \citep{tqdm}, \texttt{CASA} \citep{CASA}, \texttt{mpol} \citep{mpol}, \texttt{visread} \citep{visread}

\newpage
\bibliography{references}

\begin{thebibliography}{54}
\providecommand{\natexlab}[1]{#1}
\providecommand{\url}[1]{\texttt{#1}}
\expandafter\ifx\csname urlstyle\endcsname\relax
  \providecommand{\doi}[1]{doi: #1}\else
  \providecommand{\doi}{doi: \begingroup \urlstyle{rm}\Url}\fi

\bibitem[{ALMA Partnership} et~al.(2015){ALMA Partnership}, {Brogan}, {P{\'e}rez}, {Hunter}, {Dent}, {Hales}, {Hills}, {Corder}, {Fomalont}, {Vlahakis}, {Asaki}, {Barkats}, {Hirota}, {Hodge}, {Impellizzeri}, {Kneissl}, {Liuzzo}, {Lucas}, {Marcelino}, {Matsushita}, {Nakanishi}, {Phillips}, {Richards}, {Toledo}, {Aladro}, {Broguiere}, {Cortes}, {Cortes}, {Espada}, {Galarza}, {Garcia-Appadoo}, {Guzman-Ramirez}, {Humphreys}, {Jung}, {Kameno}, {Laing}, {Leon}, {Marconi}, {Mignano}, {Nikolic}, {Nyman}, {Radiszcz}, {Remijan}, {Rod{\'o}n}, {Sawada}, {Takahashi}, {Tilanus}, {Vila Vilaro}, {Watson}, {Wiklind}, {Akiyama}, {Chapillon}, {de Gregorio-Monsalvo}, {Di Francesco}, {Gueth}, {Kawamura}, {Lee}, {Nguyen Luong}, {Mangum}, {Pietu}, {Sanhueza}, {Saigo}, {Takakuwa}, {Ubach}, {van Kempen}, {Wootten}, {Castro-Carrizo}, {Francke}, {Gallardo}, {Garcia}, {Gonzalez}, {Hill}, {Kaminski}, {Kurono}, {Liu}, {Lopez}, {Morales}, {Plarre}, {Schieven}, {Testi}, {Videla}, {Villard}, {Andreani}, {Hibbard}, and {Tatematsu}]{HTLau2014}
{ALMA Partnership}, C.~L. {Brogan}, L.~M. {P{\'e}rez}, T.~R. {Hunter}, W.~R.~F. {Dent}, A.~S. {Hales}, R.~E. {Hills}, S.~{Corder}, E.~B. {Fomalont}, C.~{Vlahakis}, Y.~{Asaki}, D.~{Barkats}, A.~{Hirota}, J.~A. {Hodge}, C.~M.~V. {Impellizzeri}, R.~{Kneissl}, E.~{Liuzzo}, R.~{Lucas}, N.~{Marcelino}, S.~{Matsushita}, K.~{Nakanishi}, N.~{Phillips}, A.~M.~S. {Richards}, I.~{Toledo}, R.~{Aladro}, D.~{Broguiere}, J.~R. {Cortes}, P.~C. {Cortes}, D.~{Espada}, F.~{Galarza}, D.~{Garcia-Appadoo}, L.~{Guzman-Ramirez}, E.~M. {Humphreys}, T.~{Jung}, S.~{Kameno}, R.~A. {Laing}, S.~{Leon}, G.~{Marconi}, A.~{Mignano}, B.~{Nikolic}, L.~A. {Nyman}, M.~{Radiszcz}, A.~{Remijan}, J.~A. {Rod{\'o}n}, T.~{Sawada}, S.~{Takahashi}, R.~P.~J. {Tilanus}, B.~{Vila Vilaro}, L.~C. {Watson}, T.~{Wiklind}, E.~{Akiyama}, E.~{Chapillon}, I.~{de Gregorio-Monsalvo}, J.~{Di Francesco}, F.~{Gueth}, A.~{Kawamura}, C.~F. {Lee}, Q.~{Nguyen Luong}, J.~{Mangum}, V.~{Pietu}, P.~{Sanhueza}, K.~{Saigo}, S.~{Takakuwa}, C.~{Ubach}, T.~{van Kempen},
  A.~{Wootten}, A.~{Castro-Carrizo}, H.~{Francke}, J.~{Gallardo}, J.~{Garcia}, S.~{Gonzalez}, T.~{Hill}, T.~{Kaminski}, Y.~{Kurono}, H.~Y. {Liu}, C.~{Lopez}, F.~{Morales}, K.~{Plarre}, G.~{Schieven}, L.~{Testi}, L.~{Videla}, E.~{Villard}, P.~{Andreani}, J.~E. {Hibbard}, and K.~{Tatematsu}.
\newblock {The 2014 ALMA Long Baseline Campaign: First Results from High Angular Resolution Observations toward the HL Tau Region}.
\newblock \emph{\apjl}, 808\penalty0 (1):\penalty0 L3, July 2015.
\newblock \doi{10.1088/2041-8205/808/1/L3}.

\bibitem[Laporte et~al.(2021)Laporte, Meyer, Ellis, Robertson, Chisholm, and Roberts-Borsani]{Laporte2021}
N~Laporte, R~A Meyer, R~S Ellis, B~E Robertson, J~Chisholm, and G~W Roberts-Borsani.
\newblock {Probing cosmic dawn: Ages and star formation histories of candidate z $\geq$ 9 galaxies}.
\newblock \emph{Monthly Notices of the Royal Astronomical Society}, 505\penalty0 (3):\penalty0 3336--3346, 06 2021.
\newblock ISSN 0035-8711.
\newblock \doi{10.1093/mnras/stab1239}.
\newblock URL \url{https://doi.org/10.1093/mnras/stab1239}.

\bibitem[H{\"o}gbom(1974)]{hogbom1974aperture}
JA~H{\"o}gbom.
\newblock Aperture synthesis with a non-regular distribution of interferometer baselines.
\newblock \emph{Astronomy and Astrophysics Supplement, Vol. 15, p. 417}, 15:\penalty0 417, 1974.

\bibitem[Cornwell(2008)]{cornwell2008multiscale}
Tim~J Cornwell.
\newblock Multiscale clean deconvolution of radio synthesis images.
\newblock \emph{IEEE Journal of selected topics in signal processing}, 2\penalty0 (5):\penalty0 793--801, 2008.

\bibitem[Rau and Cornwell(2011)]{rau2011multi}
Urvashi Rau and Tim~J Cornwell.
\newblock A multi-scale multi-frequency deconvolution algorithm for synthesis imaging in radio interferometry.
\newblock \emph{Astronomy \& Astrophysics}, 532:\penalty0 A71, 2011.

\bibitem[Offringa and Smirnov(2017)]{offringa2017optimized}
AR~Offringa and O~Smirnov.
\newblock An optimized algorithm for multiscale wideband deconvolution of radio astronomical images.
\newblock \emph{Monthly Notices of the Royal Astronomical Society}, 471\penalty0 (1):\penalty0 301--316, 2017.

\bibitem[{Cornwell}(2009)]{2009_clean_review}
T.~J. {Cornwell}.
\newblock {Hogbom's CLEAN algorithm. Impact on astronomy and beyond. Commentary on: H{\"o}gbom J. A., 1974, A\&AS, 15, 417}.
\newblock \emph{\aap}, 500\penalty0 (1):\penalty0 65--66, June 2009.
\newblock \doi{10.1051/0004-6361/200912148}.

\bibitem[{Arras} et~al.(2021){Arras}, {Bester}, {Perley}, {Leike}, {Smirnov}, {Westermann}, and {En{\ss}lin}]{2021_bayesian_vs_clean}
Philipp {Arras}, Hertzog~L. {Bester}, Richard~A. {Perley}, Reimar {Leike}, Oleg {Smirnov}, R{\"u}diger {Westermann}, and Torsten~A. {En{\ss}lin}.
\newblock {Comparison of classical and Bayesian imaging in radio interferometry. Cygnus A with CLEAN and resolve}.
\newblock \emph{\aap}, 646:\penalty0 A84, February 2021.
\newblock \doi{10.1051/0004-6361/202039258}.

\bibitem[Czekala et~al.(2021{\natexlab{a}})Czekala, Zawadzki, Loomis, Grzybowski, Frazier, and Quinn]{mpol}
Ian Czekala, Brianna Zawadzki, Ryan Loomis, Hannah Grzybowski, Robert Frazier, and Tyler Quinn.
\newblock Mpol-dev/mpol: v0.1.1 release, June 2021{\natexlab{a}}.
\newblock URL \url{https://doi.org/10.5281/zenodo.4939048}.

\bibitem[{Junklewitz} et~al.(2016){Junklewitz}, {Bell}, {Selig}, and {En{\ss}lin}]{2016A&A...586A..76J}
H.~{Junklewitz}, M.~R. {Bell}, M.~{Selig}, and T.~A. {En{\ss}lin}.
\newblock {RESOLVE: A new algorithm for aperture synthesis imaging of extended emission in radio astronomy}.
\newblock \emph{\aap}, 586:\penalty0 A76, February 2016.
\newblock \doi{10.1051/0004-6361/201323094}.

\bibitem[{Arras} et~al.(2019){Arras}, {Frank}, {Leike}, {Westermann}, and {En{\ss}lin}]{2019A&A...627A.134A}
Philipp {Arras}, Philipp {Frank}, Reimar {Leike}, R{\"u}diger {Westermann}, and Torsten~A. {En{\ss}lin}.
\newblock {Unified radio interferometric calibration and imaging with joint uncertainty quantification}.
\newblock \emph{\aap}, 627:\penalty0 A134, July 2019.
\newblock \doi{10.1051/0004-6361/201935555}.

\bibitem[Arras et~al.(2018)Arras, Knollrnüller, Junklewitz, and Enßlin]{8553533}
Philipp Arras, Jakob Knollrnüller, Henrik Junklewitz, and Torsten~A. Enßlin.
\newblock Radio imaging with information field theory.
\newblock In \emph{2018 26th European Signal Processing Conference (EUSIPCO)}, pages 2683--2687, 2018.
\newblock \doi{10.23919/EUSIPCO.2018.8553533}.

\bibitem[Knollmüller and Enßlin(2020)]{knollmüller2020metric}
Jakob Knollmüller and Torsten~A. Enßlin.
\newblock Metric gaussian variational inference.
\newblock 2020.

\bibitem[Ho et~al.(2020)Ho, Jain, and Abbeel]{Ho2020}
Jonathan Ho, Ajay Jain, and Pieter Abbeel.
\newblock Denoising diffusion probabilistic models.
\newblock 33:\penalty0 6840--6851, 2020.
\newblock URL \url{https://proceedings.neurips.cc/paper/2020/file/4c5bcfec8584af0d967f1ab10179ca 4b-Paper.pdf}.

\bibitem[Song et~al.(2021)Song, Sohl-Dickstein, Kingma, Kumar, Ermon, and Poole]{Song2021}
Yang Song, Jascha Sohl-Dickstein, Diederik~P Kingma, Abhishek Kumar, Stefano Ermon, and Ben Poole.
\newblock Score-based generative modeling through stochastic differential equations.
\newblock 2021.
\newblock URL \url{https://openreview.net/forum?id=PxTIG12RRHS}.

\bibitem[Graikos et~al.(2022)Graikos, Malkin, Jojic, and Samaras]{graikos2022diffusion}
Alexandros Graikos, Nikolay Malkin, Nebojsa Jojic, and Dimitris Samaras.
\newblock Diffusion models as plug-and-play priors.
\newblock In Alice~H. Oh, Alekh Agarwal, Danielle Belgrave, and Kyunghyun Cho, editors, \emph{Advances in Neural Information Processing Systems}, 2022.
\newblock URL \url{https://openreview.net/forum?id=yhlMZ3iR7Pu}.

\bibitem[{Andrews} et~al.(2018){Andrews}, {Huang}, {P{\'e}rez}, {Isella}, {Dullemond}, {Kurtovic}, {Guzm{\'a}n}, {Carpenter}, {Wilner}, {Zhang}, {Zhu}, {Birnstiel}, {Bai}, {Benisty}, {Hughes}, {{\"O}berg}, and {Ricci}]{Andrews2018DSHARP}
Sean~M. {Andrews}, Jane {Huang}, Laura~M. {P{\'e}rez}, Andrea {Isella}, Cornelis~P. {Dullemond}, Nicol{\'a}s~T. {Kurtovic}, Viviana~V. {Guzm{\'a}n}, John~M. {Carpenter}, David~J. {Wilner}, Shangjia {Zhang}, Zhaohuan {Zhu}, Tilman {Birnstiel}, Xue-Ning {Bai}, Myriam {Benisty}, A.~Meredith {Hughes}, Karin~I. {{\"O}berg}, and Luca {Ricci}.
\newblock {The Disk Substructures at High Angular Resolution Project (DSHARP). I. Motivation, Sample, Calibration, and Overview}.
\newblock \emph{\apjl}, 869\penalty0 (2):\penalty0 L41, December 2018.
\newblock \doi{10.3847/2041-8213/aaf741}.

\bibitem[Benítez-Llambay and Masset(2016)]{FARGO3D}
Pablo Benítez-Llambay and Frédéric~S. Masset.
\newblock Fargo3d: A new gpu-oriented mhd code.
\newblock \emph{The Astrophysical Journal Supplement Series}, 223\penalty0 (1):\penalty0 11, 2016.
\newblock URL \url{http://stacks.iop.org/0067-0049/223/i=1/a=11}.

\bibitem[{van Cittert}(1934)]{vancittert1934}
P.H {van Cittert}.
\newblock Die wahrscheinliche schwingungsverteilung in einer von einer lichtquelle direkt oder mittels einer linse beleuchteten ebene.
\newblock \emph{Physica}, 1\penalty0 (1):\penalty0 201--210, 1934.
\newblock ISSN 0031-8914.
\newblock \doi{https://doi.org/10.1016/S0031-8914(34)90026-4}.
\newblock URL \url{https://www.sciencedirect.com/science/article/pii/S0031891434900264}.

\bibitem[Zernike(1938)]{zernike1938}
F.~Zernike.
\newblock The concept of degree of coherence and its application to optical problems.
\newblock \emph{Physica}, 5\penalty0 (8):\penalty0 785--795, 1938.
\newblock ISSN 0031-8914.
\newblock \doi{https://doi.org/10.1016/S0031-8914(38)80203-2}.
\newblock URL \url{https://www.sciencedirect.com/science/article/pii/S0031891438802032}.

\bibitem[{Feng} et~al.(2023){Feng}, {Smith}, {Rubinstein}, {Chang}, {Bouman}, and {Freeman}]{Feng2023a}
Berthy~T. {Feng}, Jamie {Smith}, Michael {Rubinstein}, Huiwen {Chang}, Katherine~L. {Bouman}, and William~T. {Freeman}.
\newblock {Score-Based Diffusion Models as Principled Priors for Inverse Imaging}.
\newblock \emph{arXiv e-prints}, art. arXiv:2304.11751, April 2023.
\newblock \doi{10.48550/arXiv.2304.11751}.

\bibitem[Hyv{{\"a}}rinen(2005)]{Hyvarinen2005}
Aapo Hyv{{\"a}}rinen.
\newblock Estimation of non-normalized statistical models by score matching.
\newblock \emph{Journal of Machine Learning Research}, 6\penalty0 (24):\penalty0 695--709, 2005.
\newblock URL \url{http://jmlr.org/papers/v6/hyvarinen05a.html}.

\bibitem[Vincent(2011)]{Vincent2011}
Pascal Vincent.
\newblock A connection between score matching and denoising autoencoders.
\newblock \emph{Neural Comput.}, 23\penalty0 (7):\penalty0 1661--1674, 2011.
\newblock \doi{10.1162/NECO\_a\_00142}.
\newblock URL \url{https://doi.org/10.1162/NECO\_a\_00142}.

\bibitem[Alain and Bengio(2014)]{Alain2014}
Guillaume Alain and Yoshua Bengio.
\newblock What regularized auto-encoders learn from the data-generating distribution.
\newblock \emph{J. Mach. Learn. Res.}, 15\penalty0 (1):\penalty0 3563–3593, jan 2014.
\newblock ISSN 1532-4435.

\bibitem[{Ronneberger} et~al.(2015){Ronneberger}, {Fischer}, and {Brox}]{Ronneberger2015}
Olaf {Ronneberger}, Philipp {Fischer}, and Thomas {Brox}.
\newblock {U-Net: Convolutional Networks for Biomedical Image Segmentation}.
\newblock \emph{arXiv e-prints}, art. arXiv:1505.04597, May 2015.
\newblock \doi{10.48550/arXiv.1505.04597}.

\bibitem[{Song} et~al.(2021){Song}, {Durkan}, {Murray}, and {Ermon}]{Song2022ml}
Yang {Song}, Conor {Durkan}, Iain {Murray}, and Stefano {Ermon}.
\newblock {Maximum Likelihood Training of Score-Based Diffusion Models}.
\newblock \emph{arXiv e-prints}, art. arXiv:2101.09258, January 2021.
\newblock \doi{10.48550/arXiv.2101.09258}.

\bibitem[{Karras} et~al.(2022){Karras}, {Aittala}, {Aila}, and {Laine}]{Karras2022}
Tero {Karras}, Miika {Aittala}, Timo {Aila}, and Samuli {Laine}.
\newblock {Elucidating the Design Space of Diffusion-Based Generative Models}.
\newblock \emph{arXiv e-prints}, art. arXiv:2206.00364, June 2022.
\newblock \doi{10.48550/arXiv.2206.00364}.

\bibitem[Anderson(1982)]{Anderson1982}
Brian~D.O. Anderson.
\newblock Reverse-time diffusion equation models.
\newblock \emph{Stochastic Processes and their Applications}, 12\penalty0 (3):\penalty0 313--326, 1982.
\newblock ISSN 0304-4149.
\newblock \doi{https://doi.org/10.1016/0304-4149(82)90051-5}.
\newblock URL \url{https://www.sciencedirect.com/science/article/pii/0304414982900515}.

\bibitem[{Chung} et~al.(2022){Chung}, {Kim}, {Mccann}, {Klasky}, and {Ye}]{Chung2022}
Hyungjin {Chung}, Jeongsol {Kim}, Michael~T. {Mccann}, Marc~L. {Klasky}, and Jong~Chul {Ye}.
\newblock {Diffusion Posterior Sampling for General Noisy Inverse Problems}.
\newblock \emph{arXiv e-prints}, art. arXiv:2209.14687, September 2022.
\newblock \doi{10.48550/arXiv.2209.14687}.

\bibitem[{Feng} and {Bouman}(2023)]{Feng2023b}
Berthy~T. {Feng} and Katherine~L. {Bouman}.
\newblock {Efficient Bayesian Computational Imaging with a Surrogate Score-Based Prior}.
\newblock \emph{arXiv e-prints}, art. arXiv:2309.01949, September 2023.
\newblock \doi{10.48550/arXiv.2309.01949}.

\bibitem[{Remy} et~al.(2022){Remy}, {Lanusse}, {Jeffrey}, {Liu}, {Starck}, {Osato}, and {Schrabback}]{Remy2022}
Benjamin {Remy}, Francois {Lanusse}, Niall {Jeffrey}, Jia {Liu}, Jean-Luc {Starck}, Ken {Osato}, and Tim {Schrabback}.
\newblock {Probabilistic Mass Mapping with Neural Score Estimation}.
\newblock \emph{arXiv e-prints}, art. arXiv:2201.05561, January 2022.

\bibitem[Adam et~al.(2022)Adam, Coogan, Malkin, Legin, Perreault-Levasseur, Hezaveh, and Bengio]{Adam2022PosteriorSO}
Alexandre Adam, Adam Coogan, Nikolay Malkin, Ronan Legin, Laurence Perreault-Levasseur, Yashar~D. Hezaveh, and Yoshua Bengio.
\newblock Posterior samples of source galaxies in strong gravitational lenses with score-based priors.
\newblock \emph{ArXiv}, abs/2211.03812, 2022.
\newblock URL \url{https://api.semanticscholar.org/CorpusID:253397812}.

\bibitem[{Stone} and {Courteau}(2019)]{Stone2019}
Connor {Stone} and St{\'e}phane {Courteau}.
\newblock {The Intrinsic Scatter of the Radial Acceleration Relation}.
\newblock \emph{The Astrophysical Journal}, 882\penalty0 (1):\penalty0 6, September 2019.
\newblock \doi{10.3847/1538-4357/ab3126}.

\bibitem[{Stone} et~al.(2021){Stone}, {Courteau}, and {Arora}]{Stone2021}
Connor {Stone}, St{\'e}phane {Courteau}, and Nikhil {Arora}.
\newblock {The Intrinsic Scatter of Galaxy Scaling Relations}.
\newblock \emph{The Astrophysical Journal}, 912\penalty0 (1):\penalty0 41, May 2021.
\newblock \doi{10.3847/1538-4357/abebe4}.

\bibitem[Smith et~al.(2022)Smith, Geach, Jackson, Arora, Stone, and Courteau]{Smith2022}
Michael~J Smith, James~E Geach, Ryan~A Jackson, Nikhil Arora, Connor Stone, and Stéphane Courteau.
\newblock {Realistic galaxy image simulation via score-based generative models}.
\newblock \emph{Monthly Notices of the Royal Astronomical Society}, 511\penalty0 (2):\penalty0 1808--1818, 01 2022.
\newblock ISSN 0035-8711.
\newblock \doi{10.1093/mnras/stac130}.
\newblock URL \url{https://doi.org/10.1093/mnras/stac130}.

\bibitem[{Bottrell} et~al.(2023){Bottrell}, {Yesuf}, {Popping}, {Omori}, {Tang}, {Ding}, {Pillepich}, {Nelson}, {Eisert}, {Gao}, {Goulding}, {Kalita}, {Luo}, {Greene}, {Shi}, and {Silverman}]{Bottrell2023}
Connor {Bottrell}, Hassen~M. {Yesuf}, Gerg{\"o} {Popping}, Kiyoaki~Christopher {Omori}, Shenli {Tang}, Xuheng {Ding}, Annalisa {Pillepich}, Dylan {Nelson}, Lukas {Eisert}, Hua {Gao}, Andy~D. {Goulding}, Boris~S. {Kalita}, Wentao {Luo}, Jenny~E. {Greene}, Jingjing {Shi}, and John~D. {Silverman}.
\newblock {IllustrisTNG in the HSC-SSP: image data release and the major role of mini mergers as drivers of asymmetry and star formation}.
\newblock \emph{arXiv e-prints}, art. arXiv:2308.14793, August 2023.
\newblock \doi{10.48550/arXiv.2308.14793}.

\bibitem[Camps and Baes(2020)]{Camps2020}
P.~Camps and M.~Baes.
\newblock Skirt 9: Redesigning an advanced dust radiative transfer code to allow kinematics, line transfer and polarization by aligned dust grains.
\newblock \emph{Astronomy and Computing}, 31:\penalty0 100381, 2020.
\newblock ISSN 2213-1337.
\newblock \doi{https://doi.org/10.1016/j.ascom.2020.100381}.
\newblock URL \url{https://www.sciencedirect.com/science/article/pii/S2213133720300354}.

\bibitem[{Nelson} et~al.(2019){Nelson}, {Springel}, {Pillepich}, {Rodriguez-Gomez}, {Torrey}, {Genel}, {Vogelsberger}, {Pakmor}, {Marinacci}, {Weinberger}, {Kelley}, {Lovell}, {Diemer}, and {Hernquist}]{Nelson2019}
Dylan {Nelson}, Volker {Springel}, Annalisa {Pillepich}, Vicente {Rodriguez-Gomez}, Paul {Torrey}, Shy {Genel}, Mark {Vogelsberger}, Ruediger {Pakmor}, Federico {Marinacci}, Rainer {Weinberger}, Luke {Kelley}, Mark {Lovell}, Benedikt {Diemer}, and Lars {Hernquist}.
\newblock {The IllustrisTNG simulations: public data release}.
\newblock \emph{Computational Astrophysics and Cosmology}, 6\penalty0 (1):\penalty0 2, May 2019.
\newblock \doi{10.1186/s40668-019-0028-x}.

\bibitem[{Weinberger} et~al.(2017){Weinberger}, {Springel}, {Hernquist}, {Pillepich}, {Marinacci}, {Pakmor}, {Nelson}, {Genel}, {Vogelsberger}, {Naiman}, and {Torrey}]{Weinberger2017}
Rainer {Weinberger}, Volker {Springel}, Lars {Hernquist}, Annalisa {Pillepich}, Federico {Marinacci}, R{\"u}diger {Pakmor}, Dylan {Nelson}, Shy {Genel}, Mark {Vogelsberger}, Jill {Naiman}, and Paul {Torrey}.
\newblock {Simulating galaxy formation with black hole driven thermal and kinetic feedback}.
\newblock \emph{\mnras}, 465\penalty0 (3):\penalty0 3291--3308, Mar 2017.
\newblock \doi{10.1093/mnras/stw2944}.

\bibitem[{Pillepich} et~al.(2018){Pillepich}, {Springel}, {Nelson}, {Genel}, {Naiman}, {Pakmor}, {Hernquist}, {Torrey}, {Vogelsberger}, {Weinberger}, and {Marinacci}]{Pillepich2018}
A.~{Pillepich}, V.~{Springel}, D.~{Nelson}, S.~{Genel}, J.~{Naiman}, R.~{Pakmor}, L.~{Hernquist}, P.~{Torrey}, M.~{Vogelsberger}, R.~{Weinberger}, and F.~{Marinacci}.
\newblock {Simulating galaxy formation with the IllustrisTNG model}.
\newblock \emph{\mnras}, 473:\penalty0 4077--4106, January 2018.
\newblock \doi{10.1093/mnras/stx2656}.

\bibitem[Zawadzki et~al.(2023)Zawadzki, Czekala, Loomis, Quinn, Grzybowski, Frazier, Jennings, Nizam, and Jian]{Zawadzki_2023}
Brianna Zawadzki, Ian Czekala, Ryan~A. Loomis, Tyler Quinn, Hannah Grzybowski, Robert~C. Frazier, Jeff Jennings, Kadri~M. Nizam, and Yina Jian.
\newblock Regularized maximum likelihood image synthesis and validation for {ALMA} continuum observations of protoplanetary disks.
\newblock \emph{Publications of the Astronomical Society of the Pacific}, 135\penalty0 (1048):\penalty0 064503, jun 2023.
\newblock \doi{10.1088/1538-3873/acdf84}.
\newblock URL \url{https://doi.org/10.1088%2F1538-3873%2Facdf84}.

\bibitem[Cooley and Tukey(1965)]{CooleyFFT1965}
James Cooley and John Tukey.
\newblock An algorithm for the machine calculation of complex fourier series.
\newblock \emph{Mathematics of Computation}, 19\penalty0 (90):\penalty0 297--301, 1965.

\bibitem[{Kurtovic} et~al.(2018){Kurtovic}, {P{\'e}rez}, {Benisty}, {Zhu}, {Zhang}, {Huang}, {Andrews}, {Dullemond}, {Isella}, {Bai}, {Carpenter}, {Guzm{\'a}n}, {Ricci}, and {Wilner}]{2018ApJ...869L..44K}
Nicol{\'a}s~T. {Kurtovic}, Laura~M. {P{\'e}rez}, Myriam {Benisty}, Zhaohuan {Zhu}, Shangjia {Zhang}, Jane {Huang}, Sean~M. {Andrews}, Cornelis~P. {Dullemond}, Andrea {Isella}, Xue-Ning {Bai}, John~M. {Carpenter}, Viviana~V. {Guzm{\'a}n}, Luca {Ricci}, and David~J. {Wilner}.
\newblock {The Disk Substructures at High Angular Resolution Project (DSHARP). IV. Characterizing Substructures and Interactions in Disks around Multiple Star Systems}.
\newblock \emph{\apjl}, 869\penalty0 (2):\penalty0 L44, December 2018.
\newblock \doi{10.3847/2041-8213/aaf746}.

\bibitem[Zhang et~al.(2018)Zhang, Zhu, Huang, Guzm{\'{a} }n, Andrews, Birnstiel, Dullemond, Carpenter, Isella, P{\'{e}}rez, Benisty, Wilner, Baruteau, Bai, and Ricci]{Zhang_2018}
Shangjia Zhang, Zhaohuan Zhu, Jane Huang, Viviana~V. Guzm{\'{a} }n, Sean~M. Andrews, Tilman Birnstiel, Cornelis~P. Dullemond, John~M. Carpenter, Andrea Isella, Laura~M. P{\'{e}}rez, Myriam Benisty, David~J. Wilner, Cl{\'{e}}ment Baruteau, Xue-Ning Bai, and Luca Ricci.
\newblock The disk substructures at high angular resolution project ({DSHARP}). {VII}. the planet{\textendash}disk interactions interpretation.
\newblock \emph{The Astrophysical Journal}, 869\penalty0 (2):\penalty0 L47, dec 2018.
\newblock \doi{10.3847/2041-8213/aaf744}.
\newblock URL \url{https://doi.org/10.3847%2F2041-8213%2Faaf744}.

\bibitem[Lemos et~al.(2023)Lemos, Coogan, Hezaveh, and Perreault-Levasseur]{lemos2023samplingbased}
Pablo Lemos, Adam Coogan, Yashar Hezaveh, and Laurence Perreault-Levasseur.
\newblock Sampling-based accuracy testing of posterior estimators for general inference, 2023.

\bibitem[{Astropy Collaboration} et~al.(2013){Astropy Collaboration}, {Robitaille}, {Tollerud}, {Greenfield}, {Droettboom}, {Bray}, {Aldcroft}, {Davis}, {Ginsburg}, {Price-Whelan}, {Kerzendorf}, {Conley}, {Crighton}, {Barbary}, {Muna}, {Ferguson}, {Grollier}, {Parikh}, {Nair}, {Unther}, {Deil}, {Woillez}, {Conseil}, {Kramer}, {Turner}, {Singer}, {Fox}, {Weaver}, {Zabalza}, {Edwards}, {Azalee Bostroem}, {Burke}, {Casey}, {Crawford}, {Dencheva}, {Ely}, {Jenness}, {Labrie}, {Lim}, {Pierfederici}, {Pontzen}, {Ptak}, {Refsdal}, {Servillat}, and {Streicher}]{astropy:2013}
{Astropy Collaboration}, T.~P. {Robitaille}, E.~J. {Tollerud}, P.~{Greenfield}, M.~{Droettboom}, E.~{Bray}, T.~{Aldcroft}, M.~{Davis}, A.~{Ginsburg}, A.~M. {Price-Whelan}, W.~E. {Kerzendorf}, A.~{Conley}, N.~{Crighton}, K.~{Barbary}, D.~{Muna}, H.~{Ferguson}, F.~{Grollier}, M.~M. {Parikh}, P.~H. {Nair}, H.~M. {Unther}, C.~{Deil}, J.~{Woillez}, S.~{Conseil}, R.~{Kramer}, J.~E.~H. {Turner}, L.~{Singer}, R.~{Fox}, B.~A. {Weaver}, V.~{Zabalza}, Z.~I. {Edwards}, K.~{Azalee Bostroem}, D.~J. {Burke}, A.~R. {Casey}, S.~M. {Crawford}, N.~{Dencheva}, J.~{Ely}, T.~{Jenness}, K.~{Labrie}, P.~L. {Lim}, F.~{Pierfederici}, A.~{Pontzen}, A.~{Ptak}, B.~{Refsdal}, M.~{Servillat}, and O.~{Streicher}.
\newblock {Astropy: A community Python package for astronomy}.
\newblock \emph{Astronomy and Astrophysics}, 558:\penalty0 A33, October 2013.
\newblock \doi{10.1051/0004-6361/201322068}.

\bibitem[{Astropy Collaboration} et~al.(2018){Astropy Collaboration}, {Price-Whelan}, {Sip{\H{o}}cz}, {G{\"u}nther}, {Lim}, {Crawford}, {Conseil}, {Shupe}, {Craig}, {Dencheva}, {Ginsburg}, {Vand erPlas}, {Bradley}, {P{\'e}rez-Su{\'a}rez}, {de Val-Borro}, {Aldcroft}, {Cruz}, {Robitaille}, {Tollerud}, {Ardelean}, {Babej}, {Bach}, {Bachetti}, {Bakanov}, {Bamford}, {Barentsen}, {Barmby}, {Baumbach}, {Berry}, {Biscani}, {Boquien}, {Bostroem}, {Bouma}, {Brammer}, {Bray}, {Breytenbach}, {Buddelmeijer}, {Burke}, {Calderone}, {Cano Rodr{\'\i}guez}, {Cara}, {Cardoso}, {Cheedella}, {Copin}, {Corrales}, {Crichton}, {D'Avella}, {Deil}, {Depagne}, {Dietrich}, {Donath}, {Droettboom}, {Earl}, {Erben}, {Fabbro}, {Ferreira}, {Finethy}, {Fox}, {Garrison}, {Gibbons}, {Goldstein}, {Gommers}, {Greco}, {Greenfield}, {Groener}, {Grollier}, {Hagen}, {Hirst}, {Homeier}, {Horton}, {Hosseinzadeh}, {Hu}, {Hunkeler}, {Ivezi{\'c}}, {Jain}, {Jenness}, {Kanarek}, {Kendrew}, {Kern}, {Kerzendorf}, {Khvalko}, {King}, {Kirkby}, {Kulkarni},
  {Kumar}, {Lee}, {Lenz}, {Littlefair}, {Ma}, {Macleod}, {Mastropietro}, {McCully}, {Montagnac}, {Morris}, {Mueller}, {Mumford}, {Muna}, {Murphy}, {Nelson}, {Nguyen}, {Ninan}, {N{\"o}the}, {Ogaz}, {Oh}, {Parejko}, {Parley}, {Pascual}, {Patil}, {Patil}, {Plunkett}, {Prochaska}, {Rastogi}, {Reddy Janga}, {Sabater}, {Sakurikar}, {Seifert}, {Sherbert}, {Sherwood-Taylor}, {Shih}, {Sick}, {Silbiger}, {Singanamalla}, {Singer}, {Sladen}, {Sooley}, {Sornarajah}, {Streicher}, {Teuben}, {Thomas}, {Tremblay}, {Turner}, {Terr{\'o}n}, {van Kerkwijk}, {de la Vega}, {Watkins}, {Weaver}, {Whitmore}, {Woillez}, {Zabalza}, and {Astropy Contributors}]{astropy:2018}
{Astropy Collaboration}, A.~M. {Price-Whelan}, B.~M. {Sip{\H{o}}cz}, H.~M. {G{\"u}nther}, P.~L. {Lim}, S.~M. {Crawford}, S.~{Conseil}, D.~L. {Shupe}, M.~W. {Craig}, N.~{Dencheva}, A.~{Ginsburg}, J.~T. {Vand erPlas}, L.~D. {Bradley}, D.~{P{\'e}rez-Su{\'a}rez}, M.~{de Val-Borro}, T.~L. {Aldcroft}, K.~L. {Cruz}, T.~P. {Robitaille}, E.~J. {Tollerud}, C.~{Ardelean}, T.~{Babej}, Y.~P. {Bach}, M.~{Bachetti}, A.~V. {Bakanov}, S.~P. {Bamford}, G.~{Barentsen}, P.~{Barmby}, A.~{Baumbach}, K.~L. {Berry}, F.~{Biscani}, M.~{Boquien}, K.~A. {Bostroem}, L.~G. {Bouma}, G.~B. {Brammer}, E.~M. {Bray}, H.~{Breytenbach}, H.~{Buddelmeijer}, D.~J. {Burke}, G.~{Calderone}, J.~L. {Cano Rodr{\'\i}guez}, M.~{Cara}, J.~V.~M. {Cardoso}, S.~{Cheedella}, Y.~{Copin}, L.~{Corrales}, D.~{Crichton}, D.~{D'Avella}, C.~{Deil}, {\'E}.~{Depagne}, J.~P. {Dietrich}, A.~{Donath}, M.~{Droettboom}, N.~{Earl}, T.~{Erben}, S.~{Fabbro}, L.~A. {Ferreira}, T.~{Finethy}, R.~T. {Fox}, L.~H. {Garrison}, S.~L.~J. {Gibbons}, D.~A. {Goldstein}, R.~{Gommers},
  J.~P. {Greco}, P.~{Greenfield}, A.~M. {Groener}, F.~{Grollier}, A.~{Hagen}, P.~{Hirst}, D.~{Homeier}, A.~J. {Horton}, G.~{Hosseinzadeh}, L.~{Hu}, J.~S. {Hunkeler}, {\v{Z}}.~{Ivezi{\'c}}, A.~{Jain}, T.~{Jenness}, G.~{Kanarek}, S.~{Kendrew}, N.~S. {Kern}, W.~E. {Kerzendorf}, A.~{Khvalko}, J.~{King}, D.~{Kirkby}, A.~M. {Kulkarni}, A.~{Kumar}, A.~{Lee}, D.~{Lenz}, S.~P. {Littlefair}, Z.~{Ma}, D.~M. {Macleod}, M.~{Mastropietro}, C.~{McCully}, S.~{Montagnac}, B.~M. {Morris}, M.~{Mueller}, S.~J. {Mumford}, D.~{Muna}, N.~A. {Murphy}, S.~{Nelson}, G.~H. {Nguyen}, J.~P. {Ninan}, M.~{N{\"o}the}, S.~{Ogaz}, S.~{Oh}, J.~K. {Parejko}, N.~{Parley}, S.~{Pascual}, R.~{Patil}, A.~A. {Patil}, A.~L. {Plunkett}, J.~X. {Prochaska}, T.~{Rastogi}, V.~{Reddy Janga}, J.~{Sabater}, P.~{Sakurikar}, M.~{Seifert}, L.~E. {Sherbert}, H.~{Sherwood-Taylor}, A.~Y. {Shih}, J.~{Sick}, M.~T. {Silbiger}, S.~{Singanamalla}, L.~P. {Singer}, P.~H. {Sladen}, K.~A. {Sooley}, S.~{Sornarajah}, O.~{Streicher}, P.~{Teuben}, S.~W. {Thomas}, G.~R.
  {Tremblay}, J.~E.~H. {Turner}, V.~{Terr{\'o}n}, M.~H. {van Kerkwijk}, A.~{de la Vega}, L.~L. {Watkins}, B.~A. {Weaver}, J.~B. {Whitmore}, J.~{Woillez}, V.~{Zabalza}, and {Astropy Contributors}.
\newblock {The Astropy Project: Building an Open-science Project and Status of the v2.0 Core Package}.
\newblock \emph{Astronomical Journal}, 156\penalty0 (3):\penalty0 123, September 2018.
\newblock \doi{10.3847/1538-3881/aabc4f}.

\bibitem[Kluyver et~al.(2016)Kluyver, Ragan-Kelley, P{\'e}rez, Granger, Bussonnier, Frederic, Kelley, Hamrick, Grout, Corlay, Ivanov, Avila, Abdalla, Willing, and development team]{jupyter}
Thomas Kluyver, Benjamin Ragan-Kelley, Fernando P{\'e}rez, Brian Granger, Matthias Bussonnier, Jonathan Frederic, Kyle Kelley, Jessica Hamrick, Jason Grout, Sylvain Corlay, Paul Ivanov, Dami{\'a}n Avila, Safia Abdalla, Carol Willing, and Jupyter development team.
\newblock Jupyter notebooks ? a publishing format for reproducible computational workflows.
\newblock In Fernando Loizides and Birgit Scmidt, editors, \emph{Positioning and Power in Academic Publishing: Players, Agents and Agendas}, pages 87--90. IOS Press, 2016.
\newblock URL \url{https://eprints.soton.ac.uk/403913/}.

\bibitem[Hunter(2007)]{matplotlib}
J.~D. Hunter.
\newblock Matplotlib: A 2d graphics environment.
\newblock \emph{Computing in Science \& Engineering}, 9\penalty0 (3):\penalty0 90--95, 2007.
\newblock \doi{10.1109/MCSE.2007.55}.

\bibitem[Harris et~al.(2020)Harris, Millman, van~der Walt, Gommers, Virtanen, Cournapeau, Wieser, Taylor, Berg, Smith, Kern, Picus, Hoyer, van Kerkwijk, Brett, Haldane, del R{\'{i}}o, Wiebe, Peterson, G{\'{e}}rard-Marchant, Sheppard, Reddy, Weckesser, Abbasi, Gohlke, and Oliphant]{numpy}
Charles~R. Harris, K.~Jarrod Millman, St{\'{e}}fan~J. van~der Walt, Ralf Gommers, Pauli Virtanen, David Cournapeau, Eric Wieser, Julian Taylor, Sebastian Berg, Nathaniel~J. Smith, Robert Kern, Matti Picus, Stephan Hoyer, Marten~H. van Kerkwijk, Matthew Brett, Allan Haldane, Jaime~Fern{\'{a}}ndez del R{\'{i}}o, Mark Wiebe, Pearu Peterson, Pierre G{\'{e}}rard-Marchant, Kevin Sheppard, Tyler Reddy, Warren Weckesser, Hameer Abbasi, Christoph Gohlke, and Travis~E. Oliphant.
\newblock Array programming with {NumPy}.
\newblock \emph{Nature}, 585\penalty0 (7825):\penalty0 357--362, September 2020.
\newblock \doi{10.1038/s41586-020-2649-2}.
\newblock URL \url{https://doi.org/10.1038/s41586-020-2649-2}.

\bibitem[Paszke et~al.(2019)Paszke, Gross, Massa, Lerer, Bradbury, Chanan, Killeen, Lin, Gimelshein, Antiga, Desmaison, Kopf, Yang, DeVito, Raison, Tejani, Chilamkurthy, Steiner, Fang, Bai, and Chintala]{pytorch}
Adam Paszke, Sam Gross, Francisco Massa, Adam Lerer, James Bradbury, Gregory Chanan, Trevor Killeen, Zeming Lin, Natalia Gimelshein, Luca Antiga, Alban Desmaison, Andreas Kopf, Edward Yang, Zachary DeVito, Martin Raison, Alykhan Tejani, Sasank Chilamkurthy, Benoit Steiner, Lu~Fang, Junjie Bai, and Soumith Chintala.
\newblock Pytorch: An imperative style, high-performance deep learning library.
\newblock In H.~Wallach, H.~Larochelle, A.~Beygelzimer, F.~d\textquotesingle Alch\'{e}-Buc, E.~Fox, and R.~Garnett, editors, \emph{Advances in Neural Information Processing Systems 32}, pages 8024--8035. Curran Associates, Inc., 2019.
\newblock URL \url{http://papers.neurips.cc/paper/9015-pytorch-an-imperative-style-high-performance-deep-learning-library.pdf}.

\bibitem[da~Costa-Luis(2019)]{tqdm}
Casper~O. da~Costa-Luis.
\newblock `tqdm`: A fast, extensible progress meter for python and cli.
\newblock \emph{Journal of Open Source Software}, 4\penalty0 (37):\penalty0 1277, 2019.
\newblock URL \url{https://doi.org/10.21105/joss.01277}.

\bibitem[Team et~al.(2022)Team, Bean, Bhatnagar, Castro, Meyer, Emonts, Garcia, Garwood, Golap, Villalba, Harris, Hayashi, Hoskins, Hsieh, Jagannathan, Kawasaki, Keimpema, Kettenis, Lopez, Marvil, Masters, McNichols, Mehringer, Miel, Moellenbrock, Montesino, Nakazato, Ott, Petry, Pokorny, Raba, Rau, Schiebel, Schweighart, Sekhar, Shimada, Small, Steeb, Sugimoto, Suoranta, Tsutsumi, van Bemmel, Verkouter, Wells, Xiong, Szomoru, Griffith, Glendenning, and Kern]{CASA}
The~CASA Team, Ben Bean, Sanjay Bhatnagar, Sandra Castro, Jennifer~Donovan Meyer, Bjorn Emonts, Enrique Garcia, Robert Garwood, Kumar Golap, Justo~Gonzalez Villalba, Pamela Harris, Yohei Hayashi, Josh Hoskins, Mingyu Hsieh, Preshanth Jagannathan, Wataru Kawasaki, Aard Keimpema, Mark Kettenis, Jorge Lopez, Joshua Marvil, Joseph Masters, Andrew McNichols, David Mehringer, Renaud Miel, George Moellenbrock, Federico Montesino, Takeshi Nakazato, Juergen Ott, Dirk Petry, Martin Pokorny, Ryan Raba, Urvashi Rau, Darrell Schiebel, Neal Schweighart, Srikrishna Sekhar, Kazuhiko Shimada, Des Small, Jan-Willem Steeb, Kanako Sugimoto, Ville Suoranta, Takahiro Tsutsumi, Ilse~M. van Bemmel, Marjolein Verkouter, Akeem Wells, Wei Xiong, Arpad Szomoru, Morgan Griffith, Brian Glendenning, and Jeff Kern.
\newblock Casa, the common astronomy software applications for radio astronomy.
\newblock \emph{Publications of the Astronomical Society of the Pacific}, 134\penalty0 (1041):\penalty0 114501, nov 2022.
\newblock \doi{10.1088/1538-3873/ac9642}.
\newblock URL \url{https://dx.doi.org/10.1088/1538-3873/ac9642}.

\bibitem[Czekala et~al.(2021{\natexlab{b}})Czekala, Loomis, Andrews, Huang, and Rosenfeld]{visread}
Ian Czekala, Ryan Loomis, Sean Andrews, Jane Huang, and Katherine Rosenfeld.
\newblock Mpol-dev/visread, January 2021{\natexlab{b}}.
\newblock URL \url{https://doi.org/10.5281/zenodo.4432501}.

\end{thebibliography}

\newpage
\appendix
\onecolumn

\section{Convolved likelihood approximation}\label{sec:approx}

\subsection{Background}\label{sec:background}
In this appendix, we perform the derivation for the core approximation used in this work to make posterior sampling tractable, namely the convolved likelihood $p_t(\mathbf{y} \mid \mathbf{x})$ in equation \eqref{eq:convolved_likelihood}. 

In the context of continuous-time diffusion models introduced by \citep{Song2021}, we work with the stochastic process $\mathbf{X}_t(\omega): \Omega \times [0, 1] \rightarrow \mathbb{R}^n$, defined on the measurable space $(\mathbb{R}^n, \mathcal{B}(\mathbb{R}^n), \{\mathcal{F}_t\}, \mathbf{W})$ where $\mathcal{B}(\mathbb{R}^n)$ is the Borel $\sigma$-algebra associated with $\mathbb{R}^n$, $\{\mathcal{F}_{t}\}$ is a set of filtrations and $\mathbf{W}$ is the Wiener measure. We will describe the stochastic process associated with the SDE of the random variable of interest $\mathbf{X}_t$ using the perturbation kernel of the SDE, $p_t(\mathbf{x}_t \mid \mathbf{x}_0)$, where $t \in [0, 1]$ is the time index of the SDE. More specifically, we consider the Gaussian perturbation kernel that correspond to the Variance-Preserving (VP) SDE \citep{Song2021,Ho2020}
\begin{equation}
    p_t(\mathbf{x}_t \mid \mathbf{x}_0) = \mathcal{N}(\mathbf{x}_t \mid \mu(t) \mathbf{x}_0, \sigma^2(t) \bbone_{n \times n})\, .
\end{equation}
In this work, $\mathbf{x}_0$ is an image of the protoplanetary disk we wish to infer from the interferometric data. Since the kernel is Gaussian, we can express $\mathbf{x}_t$, a noisy image of the protoplanetary disk, directly in term of $\mathbf{x}_0$ and pure noise,
\begin{equation}\label{eq:reparamatrization}
    \mathbf{x}_t = \mu(t)\mathbf{x}_0 + \sigma(t) \mathbf{z}\, ,
\end{equation}
where  $\mathbf{z} \sim \mathcal{N}(0, \bbone_{n \times n})$.

Since our goal is to sample from the posterior, we specify the $t=0$ boundary condition of the SDE as the posterior distribution $p(\mathbf{x}_0 \mid \mathbf{y})$. We can construct the marginal of the posterior at any time $t$ by applying Bayes' theorem, taking the logarithm and taking the gradient with respect to $\mathbf{x}_t$. However, the quantity $\grad_{\mathbf{x}_t} \log p(\mathbf{y} \mid \mathbf{x}_t)$, the second term on the RHS of equation \eqref{eq:score_bayes}, is intractable to compute since it involves an expectation over $p(\mathbf{x}_0 \mid \mathbf{x}_t)$
\begin{equation}
    p_t(\mathbf{y} \mid \mathbf{x}_t) = \int d\mathbf{x}_0\, p(\mathbf{y} \mid \mathbf{x}_0) p(\mathbf{x}_0 \mid \mathbf{x}_t)\, .
\end{equation}

To simplify this expression, we reverse the conditional $p(\mathbf{x}_0 \mid \mathbf{x}_t)$ using Bayes' theorem to get the known perturbation kernel of the SDE
\begin{equation}
    p_t(\mathbf{y} \mid \mathbf{x}_t) = \int d\mathbf{x}_0\, p(\mathbf{y} \mid \mathbf{x}_0) p(\mathbf{x}_t \mid \mathbf{x}_0) \frac{p(\mathbf{x}_0)}{p(\mathbf{x}_t)}\, .
\end{equation}

The convolved likelihood approximation involves in treating the ratio $p(\mathbf{x}_0) / p(\mathbf{x}_t) $ as a constant or setting it to 1. The argument in favor of this simplification is that at low temperature ($t \sim 0$), the prior and the convolved prior are roughly equal, $p(\mathbf{x}_t) \approx p(\mathbf{x}_0)$. Hence, our approximation converges to the true likelihood at low temperature. At high temperature ($t \sim 1$), this approximation is much worse unless $p(\mathbf{x}_0)$ can be treated as a constant over the region where $p(\mathbf{y} \mid \mathbf{x}_0)$ contributes to the integral. In other words, the convolved likelihood approximation is more accurate when the likelihood is informative (narrow).

\subsection{Convolution with a complex Gaussian likelihood}

We now evaluate the convolution between the likelihood and the perturbation kernel
\begin{equation}\label{eq:conv_lh}
    p_t(\mathbf{y} \mid \mathbf{x}_t) \approx \int d\mathbf{x}_0\, p(\mathbf{y} \mid \mathbf{x}_0) p(\mathbf{x}_t \mid \mathbf{x}_0)\, .
\end{equation}
The likelihood is a Gaussian distribution with covariance matrix $\Gamma \in \mathbb{R}^{2 m \times 2 m}$, measured empirically as described in section \ref{sec:data_and_phys},
\begin{equation}
    p(\mathbf{y} \mid \mathbf{x}_0) = \mathcal{N}(\mathbf{y} \mid A \mathbf{x}_0, \Gamma)\, .
\end{equation}
In what follows, we assume that the physical model, $A$, is not a singular matrix.  We define ${\tilde{A} \equiv S \mathcal{F} P_{\mathrm{beam}}} \in \mathbb{C}^{m \times n}$ to be the physical model. For the construction in this appendix to work, we redefine this complex matrix into its real-valued equivalent matrix
\begin{equation}
    A \equiv \begin{pmatrix}
        \Re(\tilde{A}) & -\Im(\tilde{A}) \\
        \Im(\tilde{A}) & \phantom{-}\Re(\tilde{A})
    \end{pmatrix}
    \in \mathbb{R}^{2m \times 2n}
\end{equation}
In this construction, every complex random variable, e.g., $\tilde{\mathbf{z}} \in \mathbb{C}^n$, is reformulated as a vectorized complex random variable
\begin{equation}
    \mathbf{z} = (\Re(\tilde{\mathbf{z}}), \Im(\tilde{\mathbf{z}})) \in \mathbb{R}^{2n}
\end{equation}
Naturally, we also reformulate real random variables; i.e., $\mathbf{z} \in \mathbb{R}^n$ is mapped to $(\mathbf{z}, 0) \in \mathbb{R}^{2n}$. As such, we can use the same probability rules that apply to real random variables to the complex random variables in the derivation below.

To evaluate the convolution in equation \eqref{eq:conv_lh}, we change the variables of integration for both the likelihood and the perturbation kernel. We first recall the data generating process, equation \eqref{eq:vis}, 
$
    \mathbf{y} = A \mathbf{x}_0 + \boldsymbol{\eta}\, ,
$
where $\mathbf{y} \equiv \mathcal{V}$ are the observed or simulated visibilities. We multiply equation \eqref{eq:vis} by $\mu(t)$, then use the reparameterization in equation \eqref{eq:reparamatrization} to get
\begin{equation}\label{eq:rv_conv_lh}
    \boldsymbol{\eta}_t \equiv \mu(t) \mathbf{y} - A \mathbf{x}_t = \mu(t) \boldsymbol{\eta} - A \sigma(t) \mathbf{z}\, .
\end{equation} 
We now extract the form of the perturbation kernel from equation \eqref{eq:rv_conv_lh}. Recall that $\mathbf{z} \sim \mathcal{N}(0, \bbone_{n \times n})$ is a normally distributed random variable (see appendix \ref{sec:background}). We cast this variable in the space $\mathbb{R}^{2n}$ and, in an abuse of notation, we write
\begin{equation}
    p(\mathbf{z}) = \mathcal{N}(0, \bbone_{n \times n} \oplus \bbzero_{n \times n})\, ,
\end{equation}
where $\bbzero_{n \times n}$ is a matrix filled with zeros and
\begin{equation}
    \bbone_{n \times n} \oplus \bbzero_{n \times n} \equiv \begin{pmatrix}
        \bbone_{n \times n} & \bbzero_{n \times n} \\
        \bbzero_{n \times n} & \bbzero_{n \times n}
    \end{pmatrix} \, .
\end{equation}
Despite the fact that we introduce a singular covariance matrix to account for the Dirac delta distribution of $\Im(\mathbf{z}) = 0$, the end product of our derivation will not necessarily be singular. Using this, we can now obtain the perturbation kernel of the random variable $\boldsymbol{\eta}$
\begin{equation}
    p(\boldsymbol{\eta}_t \mid \boldsymbol{\eta}) =  \mathcal{N}(\boldsymbol{\eta}_t \mid \mu(t) \boldsymbol{\eta}, \sigma^2(t)\Sigma)\, ,
\end{equation}
where
\begin{equation}\label{eq:covariance_sigma}
    \Sigma \equiv A (\bbone_{n \times n} \oplus \bbzero_{n\times n})A^T 
    = \begin{pmatrix}
        \Re(\tilde{A})\Re(\tilde{A})^T & \Re(\tilde{A})\Im(\tilde{A})^T \\
        \Im(\tilde{A})\Re(\tilde{A})^T & \Im(\tilde{A})\Im(\tilde{A})^T
    \end{pmatrix}\, .
\end{equation}

We then rewrite equation \eqref{eq:conv_lh} in terms of $\boldsymbol{\eta}$ and $\boldsymbol{\eta}_t$. Using equation \eqref{eq:rv_conv_lh}, we can relate the kernel $p(\boldsymbol{\eta}_t \mid \boldsymbol{\eta})$ to $p(\mathbf{x}_t \mid \mathbf{x}_0)$ by making use of the change of variable formula for probability densities, we obtain
\begin{equation}
    p(\boldsymbol{\eta}_t \mid \boldsymbol{\eta}) 
        = \frac{1}{\lvert \det A \rvert }p(\mathbf{x}_t \mid \mathbf{x}_0)\, .
\end{equation}
This change of variable can only occur if $A$ is not singular.
By changing the variable of integration in equation \eqref{eq:conv_lh} to $\boldsymbol{\eta}$, we introduce another Jacobian determinant, $d \boldsymbol{\eta} = \lvert\det A\rvert d \mathbf{x}_0$. Finally, we make use of the equality $p(\mathbf{y} \mid \mathbf{x}_0) = p(\boldsymbol{\eta})$ to write
\begin{equation}
    p_t(\mathbf{y} \mid \mathbf{x}_t) \approx \lvert\det A\rvert^2\int d\boldsymbol{\eta}\, p(\boldsymbol{\eta}) p(\boldsymbol{\eta}_t \mid \boldsymbol{\eta})\, . 
\end{equation}
With the approximations made in section \eqref{sec:data_and_phys}, we have $\lvert \det A \rvert = 1$. We can thus ignore this determinant factor in what follows. 
We note that the factor $\mu(t)$ scales the mean and covariance of the likelihood
\begin{equation}
    p(\mu(t) \boldsymbol{\eta}) 
    = \mathcal{N}(0, \mu^2(t) \Gamma)\, .
\end{equation}
We obtain the distribution of $\boldsymbol{\eta}_t$ by evaluating analytically the convolution implied by the sum of random variable on the RHS of equation \eqref{eq:rv_conv_lh}
\begin{equation}\label{eq:conv_lh_final}
    p(\boldsymbol{\eta}_t) = \mathcal{N}(\boldsymbol{\eta}_t  \mid 0, \mu^2(t) \Gamma + \sigma^2(t)\Sigma)\, ,
\end{equation}
which we can rewrite as 
\begin{equation}
    p(\mathbf{y} \mid \mathbf{x}_t) \approx 
        \mathcal{N}(\mu(t) \mathbf{y} \mid A\mathbf{x}_t, \mu^2(t) \Gamma + \sigma^2(t)\Sigma)\, .
\end{equation}
In the appendix that follows, we discuss in more detail the structure of the covariance matrix $\Sigma$.

\subsection{The convolved likelihood for imaging in Fourier space}

We now discuss in more detail the structure of the covariance $\Sigma$ and show how we arrive to the expression of equation \eqref{eq:convolved_likelihood} with $\Sigma_{jk} = \frac{1}{2}\delta_{jk} + \frac{1}{2} \delta_{j0}\delta_{k0}$. 
By choosing a flat beam, $P_{\mathrm{beam}} \equiv \bbone_{n \times n}$, the structure of the covariance is mostly determined by the Fourier operator, $\mathcal{F}$. The effect of the sampling function, $S$, is only to select or remove rows from the covariance matrix. As such, we invoke this function only at the end of this section.
To make things simpler, we only consider the 1D Fourier operator in order to give an intuition of the behavior for the 2D case. This intuition will extend naturally to the 2D case. 

With these simplifications in place, we replace the physical model by the unitary Fourier operator, with elements
\begin{equation}
    \tilde{A}_{k \ell} = \frac{1}{\sqrt{n}}\omega^{k \ell}\, ,\hspace{0.5cm} k,\ell \in \{0, \dots, n-1\}\, .
\end{equation}
$\omega = e^{-2 \pi i / n}$ is the n\textsuperscript{th}-root of unity and $i \equiv \sqrt{-1}$ is the imaginary number. We start by evaluating the elements of the first bloc of the covariance $\Sigma$ (see equation \eqref{eq:covariance_sigma}). Using Euler's formula and taking the real part of $\omega$, we get
\begin{equation}
    [\Re(\tilde{A})\Re(\tilde{A})^T]_{jk} = \frac{1}{n}
    \sum_{\ell = 0}^{n-1}  \cos\left(\frac{2 \pi j \ell}{n}\right)\cos\left(\frac{2 \pi k\ell}{n}\right)\, .
\end{equation}
By using the trigonometric identity $\cos(\alpha)\cos(\beta) = \frac{1}{2}(\cos(\alpha-\beta) + \cos(\alpha+\beta))$ and the general result for the sum of a geometric series, we obtain
\begin{align}
    \nonumber
    [\Re(\tilde{A})\Re(\tilde{A})^T]_{jk}
    &= \frac{1}{2n}\sum_{\ell = 0}^{n-1} \cos\left(\frac{2 \pi \ell (j - k)}{n}\right) + \frac{1}{2n}\sum_{\ell = 0}^{n-1} \cos\left(\frac{2 \pi \ell (j + k) }{n}\right) \\
    &= 
    \begin{cases}
        1, & j = k = 0 \mod n/2\\[0.5ex]
        \frac{1}{2}, & j = k \not= 0 \mod n/2 \\[0.5ex]
        \frac{1}{2}, & j + k = n \\
        0, & \text{otherwise}
    \end{cases}\, . \label{eq:block0}
\end{align}

\begin{wrapfigure}{r}{0.35\textwidth} 
    \centering
    \vspace{-0.5cm}
    \includegraphics[scale=0.45]{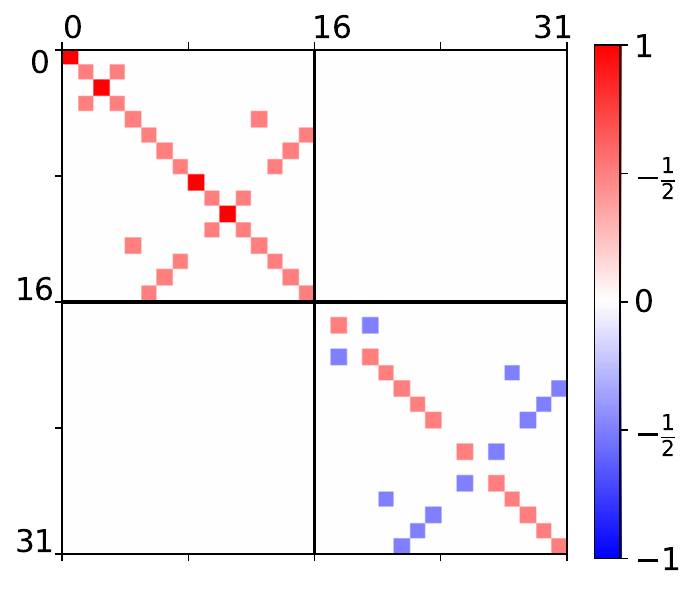}
    \caption{The covariance matrix $\Sigma$ corresponding to the 2D Fourier operator for an input in $\mathbb{R}^{(4\times 4)}$.}
    \label{fig_covariance}
    \vspace{-2.4cm}
\end{wrapfigure}

We get a formula for the other blocs of the covariance matrix using similar arguments
\begin{align}
    [\Im(\tilde{A}) \Im(\tilde{A})^T]_{jk} &= 
    \begin{cases}
        \frac{1}{2}, & j = k \not= 0 \mod n/2 \label{eq:block_1}\\
        -\frac{1}{2}, & j + k = n\\
        0, & \text{otherwise}
    \end{cases}\, ,\\
    [\Re(\tilde{A}) \Im(\tilde{A})^T]_{jk} &= 0 \, ,\\
    [\Im(\tilde{A}) \Re(\tilde{A})^T]_{jk} &= 0 \label{eq:block_4} \, .
\end{align}

In summary, we obtain a block-diagonal structure for the covariance. The diagonal components of the real and imaginary blocks are equal to $1/2$, except for their $j=k=0$ components (sometimes called the DC component) and their component at the Nyquist frequency, $j = k = n/2$. In Figure \ref{fig_covariance}, we show the covariance matrix associated to the 2D Fourier operator with $n = 4\times 4$, assuming a row-major ordering of the dimensions for an image. This covariance has a similar structure to the 1D case. As before, the diagonal components equal $1/2$ with special cases for the DC and Nyquist components (3 per block in total).  
Since our images are real-valued, the covariance is unity for the real DC component and null for the imaginary DC component. The null entries of the covariance can be removed with the sampling function since real vectors do not contribute to these Fourier components. As it turns out, the sampling function considered in this work also removes the Nyquist frequencies. This is to say that the size of the pixels considered in section \ref{sec:results} over-samples the signal in the data. As such, these cases can safely be ignored in our final expression. 

To speed up computation, we simplify further the covariance matrix by ignoring off-diagonal terms. This practical solution has little effects on the results in this work. Thus, our final expression for the covariance is $\Sigma_{jk} = \frac{1}{2}\delta_{jk} + \frac{1}{2}\delta_{j0}\delta_{k0}$, where $\delta_{jk}$ denotes the Kronecker delta.

\newpage
\section{Additional Figures}\label{app_samples}
\begin{figure}[h]
    \centering
    \includegraphics[scale=0.4]{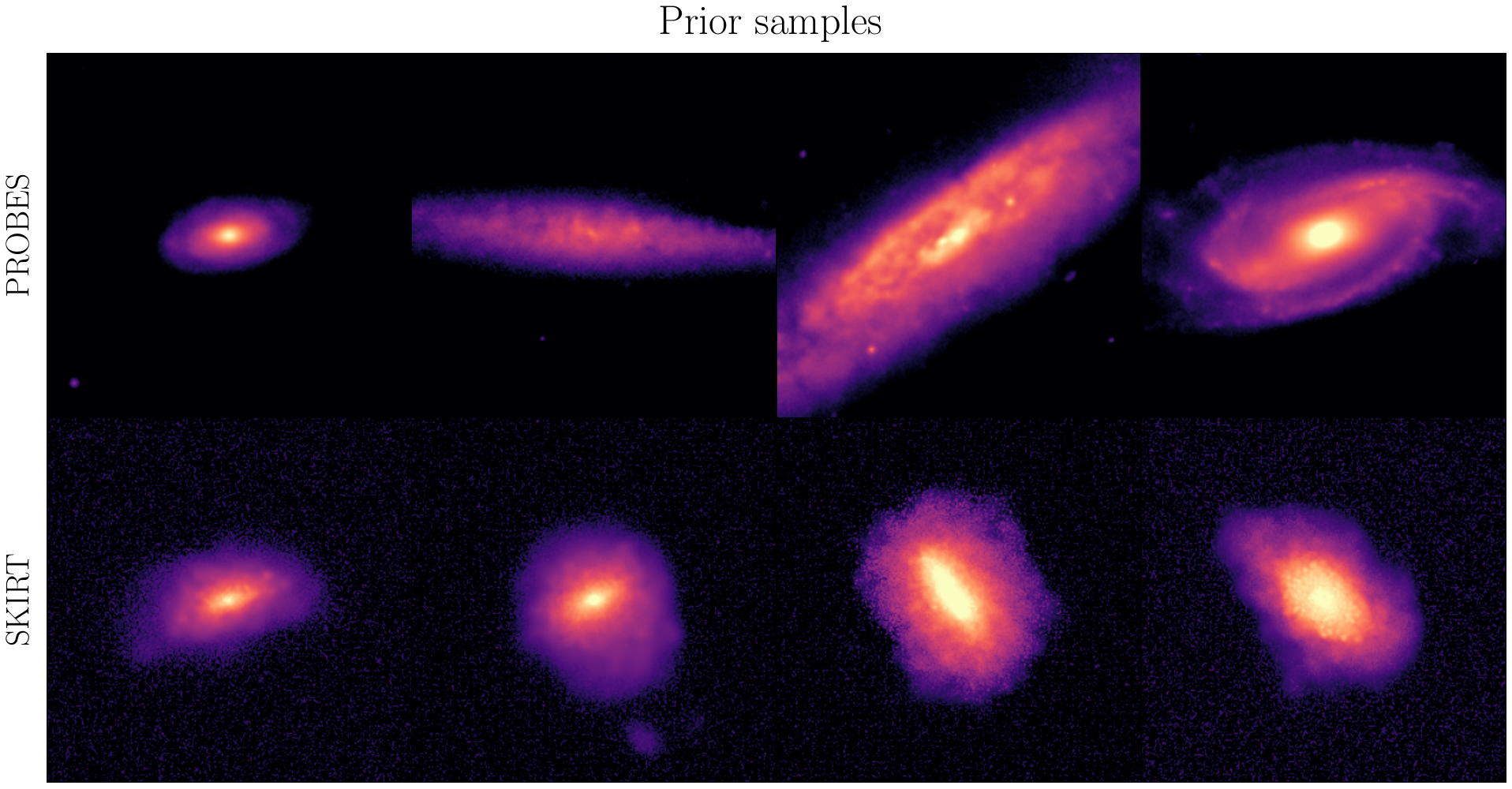}
    \caption{Prior samples from the two score-based models used as prior in this work.}
    \label{fig_prior}
\end{figure}

\begin{figure}[h]
    \centering
    \includegraphics[scale=0.4]{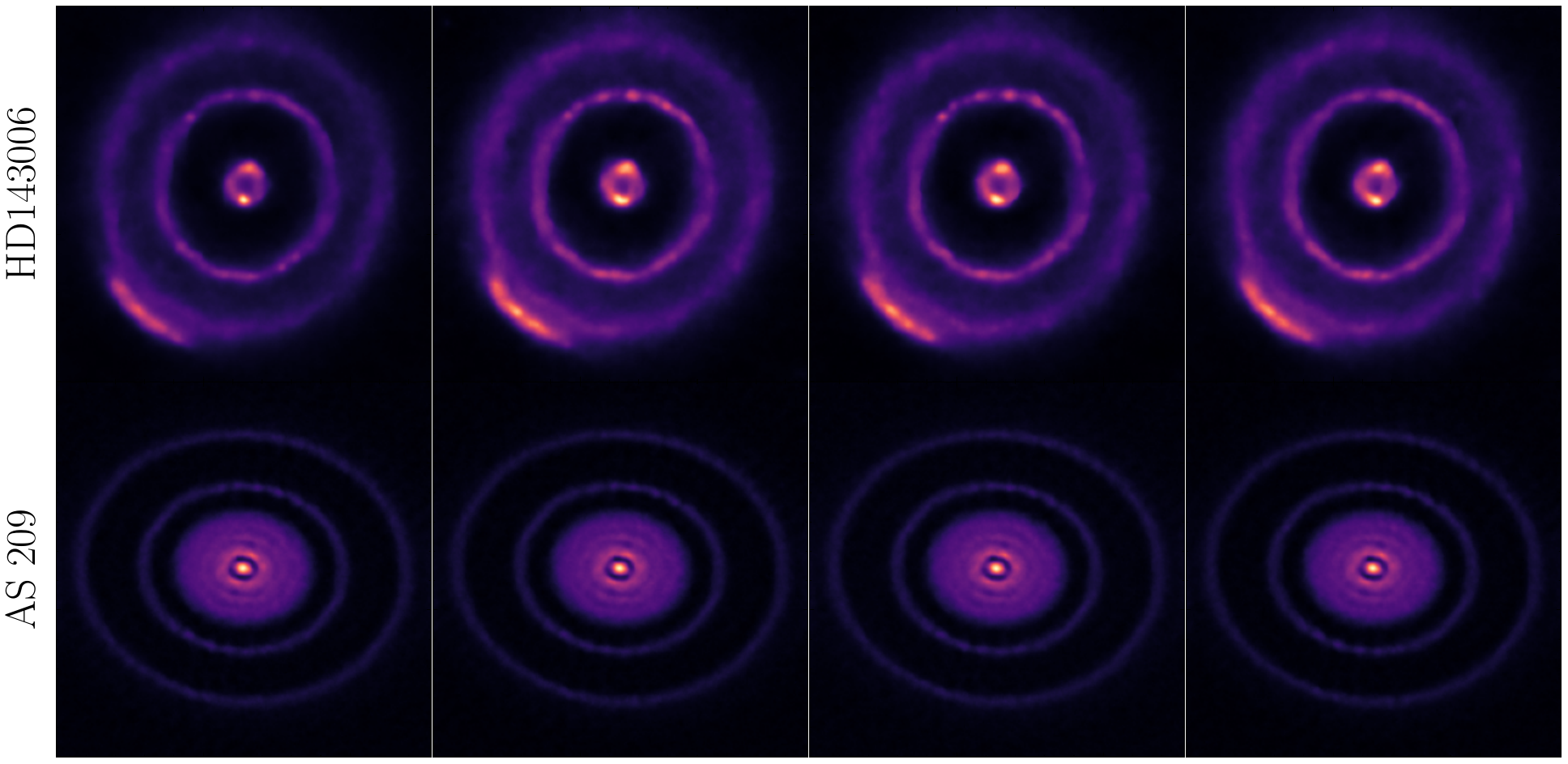}
    \caption{Posterior samples of the disks HD143006 (first row) and AS209 (second row) presented in Figure \ref{fig_res}.}
    \label{final_fig}
\end{figure}

\end{document}